# Single-molecule biophysics

Mark C. Leake[a,b]*

[a] School of Physics, Engineering and Technology, University of York, UK; [b] Department of Biology, University of York, UK

*mark.leake@york.ac.uk

Notes on Contributor:

Mark Leake holds the Anniversary Chair of Biological Physics and is the Coordinator of the Physics of Life Group at the University of York. He is also Chair of the UK Physics of Life Network (PoLNET). He heads an interdisciplinary research team in the field of single-molecule biophysics using cutting-edge biophotonics, state-of-the-art genetics, and advanced computational and theory tools. His work is highly cited, and he has won many fellowships and prizes.

# Single-molecule biophysics

Biological molecules, like all active matter, use free energy to generate force and motion which drive them out of thermal equilibrium, and undergo inherent dynamic interconversion between metastable free energy states separated by levels barely higher than stochastic thermal energy fluctuations. Here, we explore the founding and emerging approaches of the field of single-molecule biophysics which, unlike traditional ensemble average approaches, enable the detection and manipulation of individual molecules and facilitate exploration of biomolecular heterogeneity and its impact on transitional molecular kinetics and underpinning molecular interactions. We discuss the ground-breaking technological innovations which scratch far beyond the surface into open questions of real physiology, that correlate orthogonal data types and interplay empirical measurement with theoretical and computational insights, many of which are enabling artificial matter to be designed inspired by biological systems. And finally, we examine how these insights are helping to develop *new physics* framed around biology.

## 1. Introduction

Considered in isolation, a living organism appears to buck the trend of the Second Law of Thermodynamics; instead of progressing towards thermal equilibrium as defined by the maximum state of entropy there is localised ordering of the architecture of life: the tissues, cells and molecules. This out-of-equilibrium behaviour comes at the cost of a dynamic flux of free energy external to the organism which, barring some rare exceptions of microbes tapping into the chemical potential energy near hydrothermal vents and mud volcanoes and sediments which lack oxygen, comes ultimately from the light of the sun. The theme of energetic dynamics at a molecular scale is a hallmark of biological systems –like all *active* matter biological molecules use free energy to generate force and motion which drive them out of thermal equilibrium. Biological molecules are typically characterized by an inherent instability which comprises probabilistic interconversion multiple metastable free energy states which are separated by just a few multiples of $k_BT$, where $k_B$ is the Boltzmann constant and $T$ absolute temperature. In the relatively hot, wet and soft environment of active biological matter this implies complex inter-conversion kinetics. To probe this heterogeneity requires methods which can interrogate the free energy states of individual molecules, as opposed to using approaches which rely on ensemble average metrics across a large molecular population. Single-molecule biophysics is a scientific discipline whose *raison d'être* is just that, a highly interdisciplinary field that lies at the interface between not just physics and biology but percolates through the fractures crossing multiple physical and life sciences using experimental, computational and theoretical tools.

This article is structured as follows:

1. The introduction discusses the broad background of the primary physics motivation behind single-molecule biophysics of enabling exploration of the heterogeneous free energy landscape of biological molecules, as well as the primary biological motivation that physiological functions are fundamentally enabled by the regulated and energised transitioning of biological molecules between these different energy states. Details are then discussed covering the relevant physical scales of single biomolecules for length, time force and energy, and how these align with expectations based around statistical physics of the fluctuations of the water molecules which solvate biomolecules. Key historical progress involving innovations in both technology and analysis into the development of single-molecule biophysics are then discussed, in particular with the transformational use of super-resolution optical microscopy tools and the study of exemplar molecular motors. Key challenges associated with discriminating signal from noise are then outline, and the invaluable application use of high-performance computational methods in a predictive simulation capacity are then discussed in the contact of single biomolecules utilising both classical mechanics and *ab initio* quantum theory.

2. Tools which have been developed to detect and visualise single biomolecules in particular are then discussed, starting with light microscopy methods (2.1), including important details of how to pinpoint where single molecules are to a precision better than standard optical resolution limit imposed by diffraction, and how to determine if single biomolecules are interacting with each other and/or undergoing dynamic processes such as molecular turnover. The role of electrical conductance technology for single-molecule detection is then outlined, especially in the context of transformative next-generation DNA sequencing methods.

3. Methods using light to manipulate single biomolecules are then discussed, including the basics of optical tweezers, methods to manipulate and measure molecular torque and high-throughput tweezers approaches in lab-on-a-chip designs. The basics of magnetic tweezers for molecular manipulation is then outlined, including high throughput approaches, and acoustic tweezers are discussed. The use of surface probe microscopy approaches is then detailed in the context of molecular manipulation, including atomic force microscopy as well as variants of dielectric spectroscopy and electrostatic force microscopy. The role of electrical force in studying molecular torque is then detailed along with anti-Brownian electrokinetic traps.

4. The future opportunities and challenges for the field of single-molecule biophysics are then discussed. This includes speculation over increased development of correlative tools, and discussion of the statistical challenges involved in single-molecule biophysics data acquisition and analysis, and potential developments towards increased physiological relevance of experimental investigations and increased translational impacts towards areas relevant to biomedical sciences and personalised medicine.

5. The current and near future state of the field of single-molecule biophysics is the summarised, including details of the potential opportunities for developing new physics which has been inspired by biology. A glossary is also provided at the end of the article which includes details of the key acronyms used in the article.

### *1.1 Length, time, force and energy scales relevant to single biomolecules*

The processes of life share similarities to non-biological *emergent* phenomena in exhibiting relatively short and rapid length (Fig. 1) and time scale events at a molecular scale, which grow to much larger scales following subsequent interactions between molecular agents, similarly for force and energy. Single biomolecules are characterised by a nanometre length scale (nm, $10^{-9}$ m) experiencing associated forces during energised transitions, such as those in so-called molecular machines or motors discussed later, at around the piconewton level (pN, $10^{-12}$ N). Thus, the energy scale is that which can enable displacement of a nm through a pN force, or ~$10^{-21}$ J. This aligns well to that of the surrounding thermal bath of water molecules of ~$k_BT$ where $k_B$ is Boltzmann's constant and $T$ the absolute temperature; this is no coincidence, pointing towards a fundamental evolutionary adaption of, in effect, *biology to physics*. This thermal scale is that of biomolecules in motion in a cell and typical free energy differences between metastable conformations of crucial molecular machines which drive cellular processes. Cells normally energise spatially localised increases in kinetic energy of surrounding water molecules through converting the potential chemical energy locked into a high energy molecule called adenosine triphosphate (ATP) by reacting it with water to form adenosine diphosphate (ADP) with an associated release of $20k_BT$ of thermal energy per molecule; this represents a sweet spot that can fuel transitions between metastable states of molecular machines. If energy transitions involved were significantly higher there would be a risk of breaking biological chemical bonds. If the energy differences were significantly higher or lower than the probability of transitions occurring would be incredibly low due to the exponential dependence involving free energy difference $\Delta G$ in the Boltzmann factor $\exp(-\Delta G/k_BT)$; the different states cease to be *meta*stable. Such a scenario works against the needs of cellular processes which, as a rule, require responsive dynamic capabilities underpinned by frequent transitions between biomolecular states over an often rapid time scale needed in a fast-changing physicochemical cellular environment; a scale of ~$k_BT$ allows cells to hedge their bets around the challenges of their changing environment in order to maximise survival. The length scales of biological molecules is smaller than the wavelength of visible light and so biomolecules cannot be visualised directly using conventional light microscopy techniques. "Super-resolution" fluorescence microscopy can be used to achieve localisation precisions of a few tens of nanometres down to the actual nanometre length scale in some instances, whilst atomic force microscopy (AFM) can attain atomic level spatial precision equivalent to a few multiples of $10^{-10}$ m (a unit used widely throughout chemistry, atomic and solid-state physics and the structural biology research community, denoted the Ångstrom, or Å). The time scales for the processes of life cover many orders of magnitude, down in principle to femtoseconds ($10^{-15}$ s) for electron resonance time scales in the molecular orbitals of carbon-based covalent bonding, through to nanoseconds for the rotation of individual molecules in an aqueous environment, up to milliseconds, seconds and even multiple years spanning a range of mobility processes.

In optical detection, a workhouse of single-molecule biophysics, the localisation and temporal resolution depends on the number of photons collected in the signal being measured, meaning that single-molecule super-resolution imaging which might have a temporal resolution limited to around a few hundred microseconds, though more typical spatial resolution of a few tens of nanometres has an associated temporal precision of a few tens of milliseconds, capable at best of imaging molecular conformational changes which happen on a millisecond timescale, as is exploited in Förster resonance energy transfer (FRET). Fast sampling is possible with recent improvements in camera detection technologies, but this comes at the sacrifice of the finite photon emission budget and so results in a poorer spatial resolution. However, quadrant photodiodes (QPDs) used for back focal plane (BFP) detection based on laser interferometry, e.g. in optical and magnetic tweezers (discussed later), are not limited by fluorescent photon emission budgets, and indeed are also not limited by the thermal noise of electrons in the detection circuitry known as shot noise and so can record at least one thousand times faster than this (1–3).

One energy unit of $k_BT$ is equivalent to a value of around $4.1 \times 10^{-21}$ J at room temperature relevant to biology or equivalently can be denoted as 4.1 pN.nm, i.e. an energy equivalent to the work required to move against a force of around 4 pN a displacement of one nanometre.

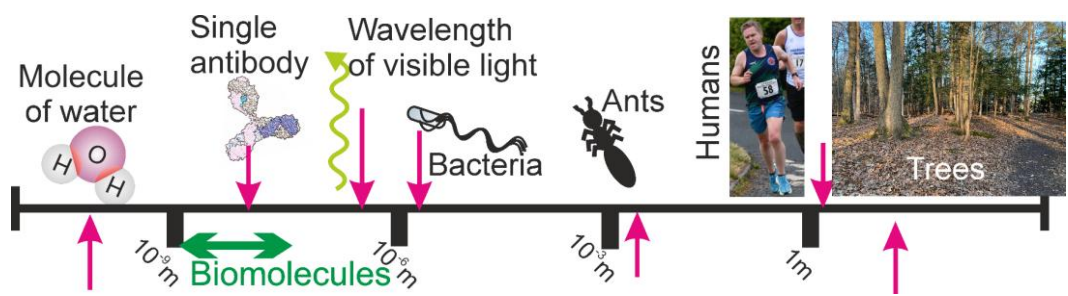

*Figure 1: Cartoon illustrating the length scale of life in the context of single biomolecules. "Humans" picture of author, credit JPO Photography.*

*1.2 Timeline single-molecule biophysics innovations*

Ensemble-average measurement techniques have produced enormous insight in the physical sciences, however, the energetic and associated conformational heterogeneity of biomolecular machinery, barring some important exceptions, results in difficulties in recapitulating the associated probability distributions for such molecular heterogeneity using typical averaging approaches due to a typical lack of synchronisation of the molecular machinery either in space or time. Well known exceptions include the study of muscle contraction (the conformational state of populations of several thousand myosin molecular motors are in effect synchronised both in time due to a chemical influx trigger of calcium and in space due to spatial periodic patterning in the myofibril filaments that makeup muscle fibres) and X-ray crystallography (several thousand copies of a specific purified biomolecule, sometimes with bound ligands present, are trapped in just a single conformation in a spatially periodic crystal). It is no coincidence that many of the formative developments in modern biophysics made from the 1950s were through structural biology using X-ray crystallography and in muscle biophysics, since for both this spatiotemporal synchronisation enables ensemble-level measurements on a population of typically several thousand molecules to yield insights in effect into single-molecule states. However, most natural physiological processes lack this degree of molecular synchronisation, which historically motivated the development of a range of single-molecule biophysics experimental techniques which can not only measure the states of single biomolecules directly but also manipulate them.

Both temporal and spatial heterogeneity can be present in a biological process, and single -molecule biophysics allows us to observe and perturb these differences. Where ensemble methods provide us with a mean average output, single-molecule biophysics approaches generate not only the average value but also renders details of the underlying probability distribution function that underlies this mean value.  Such a distribution with distinct localised maxima might indicate different metastable energetic/conformational states, and the location of the mean average relative to such maxima might indicate a preference towards one of these states, which can shift with the physicochemical environmental conditions of that biological process.

The potential physical information available from single-molecule biophysics approaches comes at a cost of technical and engineering challenges. For example, with the desire to acquire data at higher spatial resolution with faster sampling than the time scales of biology, there also needs to be a reduction in the background detection levels to achieve an effective signal-to-noise ratio greater than 1, and this is often technically non-trivial. In addition, there needs to be reliability of data collection to ensure signals acquired are associated with de facto single molecules as opposed to being due to multiples of molecules; there needs to be a reliable *molecular signature* but designing such is often challenging. A further challenge in developing single-molecule biophysics has been in enabling improved detection throughout e.g. by developing multiplexed measurement approaches, since historically detection throughput of single-molecule events has been notoriously low.

The first images of single biomolecules were reported by Hall et al in 1956, which were taken using electron microscopy (EM), an experimental technique developed from the labs of physicists. These were obtained from samples of long molecules which could form filaments, comparatively easy to identify compared to globular biomolecules, including both deoxyribonucleic acid (DNA) molecules and proteins such as collagen which is a key component of connective tissue. Carbon is a crucial element to all known forms of life, but because of its comparatively low atomic number it results in poor scattering attenuation of the electron beam in direct EM imaging. An early method developed to enhance the imaging contrast in EM was a process known as shadow casting involving a thin coating of a heavy metal such as platinum or gold evaporated and deposited onto a surface-immobilised biomolecule sample from an oblique angle. In doing so, uncoated regions of the sample can be discriminated from coated areas, which comprise a shadow-replica of the immobilised biomolecule, with relatively high contrast due to the heavy metal electron scattering. Typically, EM samples are prepared on mica due to its ability to be split into atomically flat planes; however, such planed mica exhibits a net positive surface electrical charge which potentially causes conformational artifacts on electrically charged biomolecules, such as DNA. Also, sample preparation typically involves several dehydration steps which can lead to sample distortion due to surface tension drying forces and to the structure of many biomolecules being stabilised by solvation shells.

In 1961, Boris Rotman et al were the first to report detection of biomolecules (4), albeit indirect, in a physiologically relevant water-solvated environment. Here, water droplets containing a dissolved enzyme (a biological catalyst normally, as was the case here, composed of protein) were generating at low concentrations with a signal output being fluorescence emission from the chemical reaction on the specific enzyme substrate. The droplets were nebulised onto a surface that the low concentration of enzyme and a high concentration of substrate. After allowing the enzyme-catalysed reaction to proceed for a given time the measured fluorescence from these droplets was found to be zero or an integer multiple of a given amount of fluorescence implying that the droplets with no fluorescence containing zero enzymes whereas those with integer multiples of a given amount of fluorescence signal contained one, two, three etc. enzyme molecule per droplet. The probability of the occurrence of these integer-multiple fluorescence event agreed with expectations from Poisson sampling statistics, confirming that the droplets did indeed contain a small number of enzyme molecules, and sometime as low as just one.

Following this, Thomas Hirschfeld detected single molecules of the protein globulin in water, with each molecule labelled with typically several hundred fluorescent dye molecules. Barack and Webb then reported in 1982 the tracking of single lipid molecules labelled with fluorescent dye diffusing on a cell membrane and were the first to report a method of localisation based on the locating the peak of the fluorescence intensity distribution observed on a frame by frame basis to find the centre of molecules, using nothing more advanced than latest tracing paper overlaid on subsequent photographic image frames! The significance of this low-throughput innovation was that the minimum observed spatial extent of fluorescence distribution was limited by optical diffraction to be around half a wavelength of the characteristic light emitted in fluorescence, so typically a few hundred nm, whereas pinpointing down the intensity centroid of this in effect enabled super-resolution prediction for the position of the most likely location of the fluorescent dye molecules themselves which were located in a region of space of the single molecule of only a few nanometres. Such centroid localisation and subsequent Gaussian fitting is now a key stage in many single-molecule biophysics techniques both for optical based detection and biomolecule manipulation. Gelles et al reported using a similar intensity centroid localization approach in 1988 by finding the centre of a plastic bead of around one micron in diameter ($10^{-6}$ m, 1 μm) attached to single kinesin motor molecules which caused the bead to rotate; the precision of this localisation was pushed down to a few nanometres by first digitising the bead images prior to using a Gaussian fitting algorithm to locate their centres on each image frame. Using similar methods, the first report, made in 1996, of the super-resolved localisation of a *single* fluorescent dye molecule attached to a biomolecule was performed by Schmidt et al using the dye rhodamine attached to a lipid molecule in which they were able to achieve 30 nm resolution. Following this, the next most significant innovation was report by Sako et al in 2000 who performed direct single-molecule imaging on single living cells.

*1.3 Development of transformative super-resolution microscopy*

In 2014, the Nobel Prize in Chemistry was awarded jointly to Eric Betzig, Stefan Hell and William Moerner for contributions towards establishing the field of super-resolution optical microscopy, namely imaging with a spatial resolution better than that predicted from the standard diffraction-limited optical resolution of a few hundred nanometres for normal visible light microscopy, what is now the cutting-edge of light microscopy after its inception approximately 300 years ago (5). Moerner et al were the first to report the detection of single molecules, albeit in a non-biological sample, in a solid at cryogenic temperatures, while Orrit et al used fluorescence to improve the signal-to-noise ratio of the measurement. Betzig innovated methods to develop near-field scanning microscopy (NSOM) and was the first to report associated super-resolution methods in cells, and then used these to detect single fluorescent dyes in a monolayer at room temperature. But, inside cells, fluorescent dye tags are typically far too densely packed to achieve single-molecule imaging. In the mid-1990s, Hell and Betzig independently conceived a new method to have only a single dye molecule that was actually photoactive (i.e. emitting fluorescence) in the diffraction limited excitation volume. Betzig proposed using modified fluorescent proteins and activating only a subset for each image frame; well-separated molecules could then be localised by similar intensity centroid fitting already described, before switching off their fluorescence then proceeding with another cycle of exciting a different subset of molecules until a full reconstructed image is generated, a technique denoted photoactivated localization microscopy (PALM). Hell instead developed a method which used stimulated emission depletion microscopy (STED), the theory for which was independently formulated in the 1980s by the Okhonin who patented the first super-resolution microscope based on this approach. STED reduces the volume using a donut shaped depletion bead from which a fluorescent dye molecule emits until it only contains a single molecule. By scanning this excitation volume over the sample single dye tags attached to a biomolecule can be located one at a time.

*1.4 Importance of molecular motors in the history of single-molecule biophysics*

Biological molecular machines, also called molecular motors, deserve special mention due to their crucial importance in a range of fundamental cellular processes and in the use and development of a range of different single-molecule biophysics tools to study them. They are a class of, primarily, proteins that transform chemical potential energy, typically released from ATP hydrolysis, into mechanical work through an associated molecular conformational change. These mechanical outputs underpin many biological activities such as molecular cargo transportation through to the movement of cells and the contraction of muscles. Single motors move in increments of just a few nm with associated forces of a few pN. The structures of several such motors have been studied with traditional structural biology tools such as EM, nuclear magnetic resonance and X-ray crystallography but the dynamics and motions of motors are better studied with single-molecule biophysics tools.

The molecular motor kinesin is responsible for the transportation of cell cargos, which use microtubule filaments as a molecular track. Block et al used optical tweezers (OT) to hold a plastic bead of ~1 μm diameter coated in kinesin molecules and positioned it onto microtubule. OT (discussed in more detail later in this review), creates a potential energy well in the vicinity of refractile particles using a focused laser beam. This generates optical trapping forces from a combination of light scattering and refraction which push the bead to a stable position roughly close to the laser focus. By using OT, the bead was prevented from diffusing away from the microtubule allowing stochastic interactions between the kinesin and microtubule to be explored, such as the spatial displacement due to the molecular motor stepping activities and the size of the associated forces involved. Two models had been proposed for the movement of kinesin: a stroke-release model and a hand-over-hand model, i.e. a continuous or interrupted attachment to the microtubule (Fig. 2). Block et al determined that a hand-over-hand model better explained the data. The key point is that addressing this biological question catalysed the technological improvements in single-molecule biophysics; biology *drove* a new type of physics experiment.

In a later study, Svoboda and Block further improved OT using differential interference contrast optics and dual quadrant photodiodes (QPDs) to increase the precision of measurement for spatial displacement of the trapped bead to sub-nm levels. This enhanced precision facilitated refined measurement of kinesin displacements which indicating that its velocity decreased linearly with load up to forces of 5-6 pN. They also compared the force-velocity curves at high and low ATP concentrations to deduce that the movement per catalysed ATP decreases at higher load. The bigger picture: improvements to single-molecule biophysics technology transformed our understanding of the underpinning mechanisms involved in the how molecular machines really work.

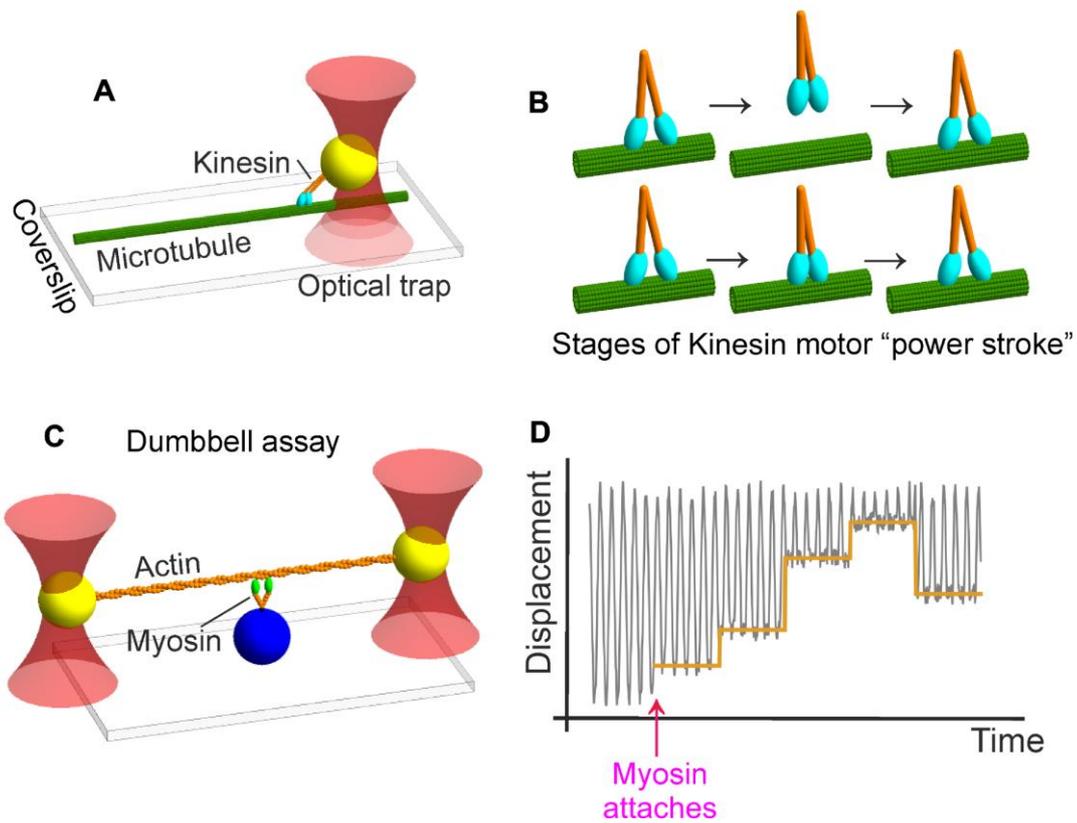

*Figure 2: Assays for molecular motor single-molecule force spectroscopy experiments. **A**. Optical tweezers position a kinesin-coated microbead onto a microtubule, applying a controllable lateral force. **B**. Two possible models of kinesin motion during its "power stroke" on a microtubule during which force is generated. Top: stroke-release model: kinesin detaches from microtubule and diffuses back later; bottom: hand-over-hand model kinesin stays attached for multiple cycles. **C**. Dumbbell assay to determine myosin motor stepping on actin filament. Here is shown a general assay; a microbead (blue) is functionalised with myosin and immobilised on a microscope coverslip. Myosin moves along actin, two synchronised optical traps detect its step size and fits force dependence as it does so. **D**. Idealised displacement of the microbead in **C**. relative to optical trap centre; initially trap position is oscillated using a triangular wave driving function. Myosin attaches to the actin and starts exerting a pulling force (magenta arrow), reducing the oscillation amplitude and allowing determination of the mean myosin displacement (yellow).*

Myosin proteins are another class of biomolecular motor, involved in transporting cell organelles along actin filaments as well as the contraction of muscles. Biochemical studies suggested that myosin-V was 'processive'; it does not fall off its actin track very frequently but rather remains in a bound state allowing it to undergo multiple "catalytic cycles", manifest as multiple "steppy" walks, before finally detaching. Mehta et al used OT to support the processive motor hypothesis and further determined its step size as ~36 nm. Fig. 2C illustrates a typical dumbbell OT configuration used; one OT has its position oscillated while the other is fixed. When a myosin molecule steps along its actin track, it pulls on the actin which is detected as a reduction in its positional noise (in a similar way to a loosely tethered washing line fluctuating more in the wind than a tightly tethered one). This detection method revealed the spatial step size, speed and direction for an individual myosin molecule in real time. Traditionally, relatively high OT forces of several tens of pN were used, but to address questions of greater physiological relevance at forces closer to the pN level itself which has motivated improvements in optical trapping stability (similarly using a complementary approach of magnetic tweezers, described later in this review).

*1.5 Challenges of noise when acquiring single-molecule signals*

A key challenge to single-molecule quantification is the small size of the various physical signals from single molecules compared to levels of noise. For example, in super-resolution fluorescence microscopy, the localisation precision of a fluorophore is a function of the number of detected photons, $N$, and is given by:

$$< (\sigma_x)^2 > = \frac{s^2 + a^2/12}{N} + \frac{4\sqrt{\pi}s^3 b^2}{aN^2} \qquad (1)$$

Here, $\sigma_x$ is the standard deviation of the spatial localisation parallel to just the $x$-axis – a measure of the overall localisation error - while $s$ is the standard deviation of the point spread function (PSF), $a$ is pixel edge length of the camera detector and $b$ is background or dark noise on these pixel measurements. To reduce the localisation error, the number $N$ of collected photons can be increased and/or the level of background noise $b$ decreased (e.g., by cooling the camera pixel array lower than room temperature either with air or water cooling). For fluorescence microscopy, a key advantage is that the emission wavelength differs from that of the excitation wavelength so an emission filter can be used to block out all but a ~50 nm window of light, hence the associated photon sampling noise is reduced due to acquiring only emitted photons which eliminates noise in the rest of the camera-sensitive spectrum, particularly the bright excitation light.

Over the past decade or more, several developments have enabled increased localisation precision in optical microscopy such that the limiting factors are the size of labelling probe and the density of probe labelling. Using only partial sample illumination is a useful trick to apply which can leave the rest of the imaging volume unilluminated and so substantially reduces out of focus fluorescence emissions. A popular example of this approach is total internal reflection fluorescence (TIRF) microscopy. In TIRF, an excitation beam propagates to a glass or quartz microscope coverslip on which a biological sample is located, at an oblique angle of incidence above the critical angle at which total internal reflection of the beam occurs. A consequence of this is that an evanescent wave is generated which propagates into the sample, decaying exponentially with distance with a characteristic exponential depth of penetration of ~100 nm; there is in effect a thin optical slice in which fluorescent dye tags will be excited, whereas beyond this there is limited excitation and so a substantial removal of "out of focal plane" fluorescence emissions, which would otherwise have been manifest on detection as a source of noise, thus increasing the effective signal-to-noise ratio for the desired signals close to the surface. A second example is that of light sheet microscopy; here a thin optical slice of a few μm illuminates the sample at any one time, which can be used for enabling single-molecule imaging in large tissue samples which otherwise would exhibit too much noise from scattering of excitation and emission light from layers of cells above and below the focal plane. A similar approach involves delimiting laser excitation to a far smaller diameter so that the excitation intensity is many fold higher, e.g. Slimfield microscopy (6) and enhanced variants which use oblique angle excitation to minimise out of focus fluorescence such as variable-angle Slimfield microscopy (SlimVar) (7), to achieve higher levels of emission while using more rapid sampling to track fast-moving biomolecules in cells. Recent improvements in developing brighter and more photostable fluorescent dye tags have also been valuable in enhancing the signal above the level of noise while denoising image algorithms have also been implemented to extract maximum signal from single biomolecule image data.

When performing force measurements on single biomolecules, associated force spectroscopy noise limits measurement precision. The source of such noise includes mechanical vibration, thermal expansion, air currents, electrical shot noise in sensors, plus Brownian noise due to the thermally driven spatial fluctuations on biomolecular position following their bombardment by water molecules at a typical frequency $\sim 10^{13}$ Hz. Force spectroscopy measurements with sub-nm resolution can be achieved in temperature-controlled ($\pm 0.2°C$), acoustically isolated measurement environments often with locating the control hardware such as lasers and computers in a separate room and/or using mechanical reinforcements on key optical mounts such to minimise vibration.

For measurement of single biomolecule displacement, simply enclosing optical components in sealed boxes results in noise reductions due to inhibition of air currents and acoustic vibration coupling. Replacing air inside such sealed boxes with helium reduces noise further; since helium has a lower density than air there are fewer particles available to scatter light. This method has been used to decrease noise spectral density 10-fold at 0.1 Hz for monitoring single protein molecule displacements with an effective noise is under 0.1 nm, which is the diameter of a single hydrogen atom.

For optical trapping experiments on single biomolecules, optical tweezers typically use near infrared (NIR) lasers as the light source (e.g. the relatively affordability Nd:YAG laser sources with wavelength 1,064 nm), which is ideal in avoiding the high-absorption region of many proteins in the visible light spectrum. However, the water absorption coefficient at 1,064 nm wavelength is $\sim 1$ cm$^{-1}$, compared to $10^{-4}$ to $10^{-2}$ cm$^{-1}$ in the visible range, and microscope slides/coverslips do not always have minimised absorption at 1,064 nm, which manifest as increased noise. In magnetic trapping, heat generation is also a potential source of noise. Despite some of this generated heat being removed with water or fan cooling, temperature gradients are still formed. Even a relatively small $\sim 1$ K gradient potentially causes mechanical drift due to thermal expansion of optical components on the order of $\sim 100$ nm; this is over an order of magnitude larger than the step size of a single kinesin molecular motor, and is also larger than the displacement of DNA-based molecular motors whose step size typically is as low as the 0.34 nm separation of individual nucleotide base pairs, and also higher than the unfolding events of modular proteins which are typically 20-30 nm. The usual way to tackle measurement error due to thermal expansion is through quantifying drift with a fiducial bead immobilised onto the microscope coverslip surface and hence correcting for such displacements.

Fig. 3 shows the typical arrangement of optical components for drift measurement and correction. Here, balanced QDPs such as those indicated can measure the position of optically trapped beads in optical tweezers in the ballistic regime of colloidal motion at MHz-GHz sampling rates. By correlating the signals from each, the common mode quantum noise can be in effect removed. For drift corrections, a green tracking laser and QPD2 is used to monitor the position of a surface-immobilised fiducial bead using laser interferometry. Its signal is then used to calculate a compensation spatial displacement of the sample, which is executed by the motorised stage. The red optical trapping laser then detects the positions of the larger "probe" bead independently by using a separate QPD (QPD1).

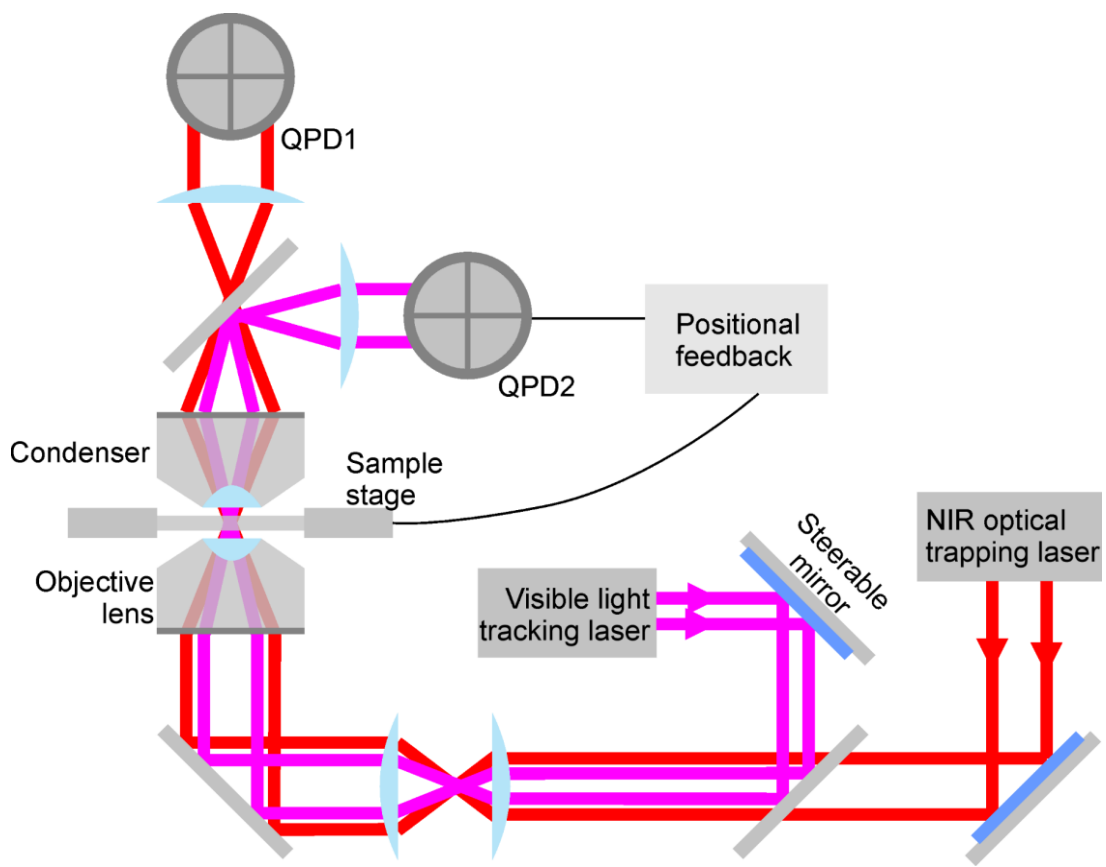

*Figure 3: Schematic of typical optical path for measuring and correcting mechanical drift of optical components. The tracking laser feeds back the position of a fiducial bead fixed on the coverslip of a sample flow cell. Any relative fluid cell movement due to mechanical drift is inputted to a feedback system that instructs a motorised nanostage to compensate for the drift, resulting ~0.1 nm stabilisation of the system along all 3 spatial axes.*

Brownian noise arises from thermal fluctuations of water molecules onto features such as the probe bead in an optical trapping experiment. Modelling the system as a damped Hookean spring and using the equipartition theorem along the x-axis indicates:

$$\frac{1}{2} k <x^2> = \frac{1}{2} k_B T \qquad (2)$$

where $x$ is the position of the bead whose mean average is calculated over a relatively long period of time of typically ~seconds compared to the ca. $1/10^{13}$ s inter-collision time of surrounding water molecules. This formulation can be used to quantify the size of the thermal noise, where $k$ is the spring constant, $k_B$ is Boltzmann constant and $T$ is the absolute temperature. Rearranging the equation indicates:

$$\Delta x = \sqrt{\frac{k_B T}{k}} \qquad (3)$$

where $\Delta x$ is the magnitude of the positional noise. Fitting the bead's positional power spectrum to a theoretical Lorentzian function enables estimation of the "corner frequency" which can be related to the trap stiffness; during this fitting procedure, the effects of other noises sources (e.g. mechanical related ~dc components) can be removed from the data by selecting a subset of the spectrum's frequency range.

*1.6 Role of computation in single-molecule biophysics*

Computational single-molecule biophysics has its origins in the 1950s, when relatively coarse simulations of the motion of hard and elastic spheres were conducted over short time scales, demonstrating the potential of molecular dynamics techniques to study dynamical properties of systems and generate molecular trajectories. With increases in computational power, higher quality data, and more realistic molecular dynamics force fields significant subsequent progress was made to allow refinement of experimental protein structures which were derived from X-ray crystallography, as well as analysis of the dynamic behaviour of interior atoms in folded proteins and larger scale protein dynamics such as simulating key sites of structural bending of a single protein (8). As key historical problem was how to best deal with simulating the thousands of water molecules necessary to solvate biomolecules in a computationally tractable way. To address this the relatively simple TIP3P model was created in the early 1980s, but due to computational limitations early studies of solvation concentrated largely on the geometry of the solvation shell around biological molecules as these include only a small number of additional residues, such as 72 in the solvation shell of a B-DNA dodecamer. It became subsequently clear that improving classical force fields would be of little benefit without a robust and computationally tractable solvation scheme which could be used in large simulations. To address this challenge, the Generalised Born solvation model was developed which implemented in a range of molecular dynamics (MD) software to improve the accuracy of simulations. Explicit water models (i.e. simulating the positions of all atoms involved) remained computationally in many cases until the development of the particle mesh Ewald technique, which enabled a cut-off in calculating the effect of solvent at long distances and so reduces computational complexity dramatically. This enhancement also facilitated fully solvated simulations of ribonucleic acid (RNA), which had previously been elusive.

Some work has been done on *ab initio* density functional theory based molecular simulations, made possible by the work on computational quantum mechanics of Car and Parrinello. However, these simulations were limited again by computational power, complexity of the system, and the sizeable memory requirements to interrogate molecular wave functions. Coarse-graining techniques to reduce the computational demand to its bare minimum were also developed during this time using classic statistical mechanics and polymer physics considerations such as the wormlike chain and freely-jointed chain, predictions about the relation between molecular force and the end-to-end extension of DNA were enabled and compared against single-molecule biophysics experiments. More computational-based models of coarse-grained molecules were developed in parallel with atomistic methods, for example grouping nearby linked atoms together as a single unit characterised by a single mass and mechanical compliance, and met with some success in enabling longer timescales to be tractably simulated, albeit accurate only under specific conditions, and it is only relatively recently that more generalised coarse-graining has been possible for nucleic acids.

A more recent development of computation has involved the application of Artificial Intelligence (AI) and Deep Learning (DL) approaches in single-molecule biophysics. A transformative development has been that of Alphafold (9) and associated variants which use DL methods trained on a wealth of experimental structural biology data to enable structural predictions from sequence data for multiple proteins whose structures have not yet been experimentally elucidated; although not a de facto single-molecule tool, the predictive power of Alphafold is impacting multiple areas of research associated with single-molecule biophysics. An area of active research involves developing improvement Alphafold predictions of nucleic acid structures.  In protein binding, the DL algorithm DeepBind is cable of predicting the sequence specificities of proteins that bind to nucleic acids, trained on *in vitro* data and tested on *in vivo* data.

AI has been successfully applied to pattern recognition of structural data in the form of cryogenic electron microscopy (Cryo-EM), in which samples are imaged at very low temperatures using rapid flash freezing to both minimise their mobility to enable unblurred imaging while also preventing the formation of detrimental ice crystals from forming in the sample, to classify images into sub-groups and thus facilitate improvements in the effective signal-to-noise ratio via averaging within each sub-group. Very recent developments have involved the development of super-resolution AI reconstructions from standard widefield fluorescence microscopy data utilising a range of different super-resolved cellular structures as gold standards for training data.

## 2. Detecting and visualising single biomolecules

Techniques which allow us to observe single biomolecules in their native state and quantify a range of their physical properties can help reveal underlying molecular interactions and complex dynamic behaviours. These experimental single-molecule biophysics approaches include several optical microscopy methods, structural investigation tools and electrical conductance measurements.

### *2.1 Optical microscopy approaches*

Several light microscopy tools have been developed for single-molecule biophysics to enable molecular detection/visualisation, many using fluorescence emission, a technique by which incident light of one wavelength is absorbed by, typically, a fluorescent dye molecule resulting in energetic elevation of a ground state electron and subsequent emission at a longer wavelength once the electron returns to its ground state due to vibrational and internal energy losses of the electron-molecule system, over a typical time scales of nanoseconds (ns i.e.$10^{-9}$ s). This separation between absorption and emission wavelengths of typically a few tens of nm enables spectral filtering to remove the excitation wavelength using dichroic mirrors made using specialised optical coatings, which dramatically increases the image contrast. Fluorescence emission also occurs at longer length scales beyond single dye molecule tags typically used, including materials such as gold, which utilises plasmon resonance effects, and fluorescence in nanodiamond and quantum dots, though the physical processes are more complex than those discussed here.

### *2.1.1 Localization microscopy to overcome the optical resolution limit*

Optical microscopy is a valuable single-molecule biophysics tool to since it is less perturbative to natural physiology than many competing methods. A spatial limit is that due to optical diffraction is the usual far-field imaging regime of optical microscopes (i.e. the camera/detector is more than a few wavelengths of light away from the light emission sources in the sample) resulting is circle of optical uncertainty is the lateral imaging plane whose radius is equivalent to roughly half a wavelength in the lateral sample plane, equivalent to the optical resolution limit. Since biomolecules are typically ~2 orders of magnitude smaller than this length scale this presents technical challenges towards pinpointing the location of any individual molecule, since when two light emitting molecules are within a few nm of each other their diffraction-limited images overlap and the positions of each cannot be accurately determined.

However, the shape of the light intensity profile distribution from a single emitter can be modelled analytically as Bessel functions (usually approximated for computational speed as a Gaussian function when recorded on pixelated photon detector devices such as cameras). Creating a low spatial density of such emitters, such that the mean nearest neighbour separation is much greater than the optical resolution limit, allows detection can find the intensity centroid through analytical fitting of any single molecule to within a few nm lateral precision (Fig. 4A). Gaussian masking, as opposed to fitting (i.e. convolution of detected fluorescent foci with a Gaussian function, Fig. 4B), sacrifices a small amount of spatial precision but is computationally faster. The intensity of detected foci can be used to estimate the number of fluorescent emitters precent inside each foci, i.e. the molecular stoichiometry (Fig. 4C), either by counting the actual number of steps or, if many (typically more than six) fluorophores are present, by using the initial intensity (10–12). Quantification using the decay rate is also possible. These techniques have been used to determine the stoichiometry of the bacterial replication complex (13), a method which has also now been extended recently to yield also dynamic information regarding how the different components of the replication molecular machinery actually turnover with respect to time (14), as well as transcription factor clusters (15–17). By extension to multi-colour microscopy, the structural maintenance of chromosome proteins used for remodelling DNA have also been investigated with these methods (18). Recently these techniques have also been combined with image deconvolution to determine the total protein copy number of a transcription factor in yeast (19).

*2.1.2 Methods involving photophysical blinking and switching*

Several methods photoactivate a subset of fluorescent dye molecules in a densely label a biological structure, such that the most likely position of any single molecule emitting fluorescence can then be pinpointed using localization methods to a precision which is better than the diffraction-limited resolution. PALM and Stochastic Optical Reconstruction Microscopy (STORM) (and easier to implement variants such as direct STORM (dSTORM)) start with all dye molecule either in a dark state for PALM or a different colour such as green in STORM, which are activated in PALM (or switched in

STORM) to a different (often red) colour by using an ultraviolet laser pulse. This results in stochastic activation/switching of a subset of dyes to an emitting state, which is then quantified using a shorter wavelength detection. Similar approaches such as Binding-Activated Localization Microscopy (BALM) and Bleaching/Blinking assisted Localization Microscopy (BaLM) can be simpler to implement as only one illumination wavelength is required and images are acquired continuously instead of stroboscopically.

Another single-molecule localization microscopy method is called PAINT (points accumulation for imaging in nanoscale topography). Here, freely diffusing fluorescent dyes bind stochastically and transiently either directly or via ligands to target biological structures; since binding is transient this achieves the same goal as stochastic photoblinking/switching to ensure that, once conditions are optimised, the mean nearest neighbour separation of dye molecules is much greater than the optical resolution limit in any given image frame. A popular variant of this approach is DNA-PAINT (20), for which a short single-stranded tag of typically 8-10 nucleotide bases acts as a docking strand on a target biological structure, with the complement image-strand which has a bound fluorescent dye tag freely diffusing in the cell. The advantage of using DNA is that the sequence can be modified to tune the binding kinetics; this enables optimisation towards generating the best super-resolution reconstruction image given a particular cell environment and target biological structure.

Another approach to increasing the mean nearest neighbour separation of fluorescent dye molecules in the sample beyond the optical resolution limit is to inflate the size the sample after it has been labelled. This approach is known as expansion microscopy (ExM). In order to achieve the expansion the sample has to be first fixed and embedded into a hydrogel, which can then be expanded my changing the hydrogel properties, typically by a factor of 4-5. The major issue with this approach is asymmetry of expansion generating artefactual sample morphologies, however, the approach is proving a valuable tool for many relatively large biological samples.

These various technologies have been extended into multiple colours, 3D, live samples, although optimising imaging buffers for multiple colour experiments and resolving clusters of fluorescent dye molecules remain significant challenges. Notable recent developments include the high lateral localisation precision of 20 nm and axially of 50 nm in a study of the 3D structure of chromatin and 3D imaging in living cells applied to study the stoichiometry of DNA polymerase PolC in live bacteria *Bacillus subtilis*.

Increase in imaging data complexity has driven advances in analysis software tools. For example, early single-molecule super-resolution fluorescence experiments utilised simple localisation software such as QuickPALM, but this has been improved using RainSTORM, ThunderSTORM, and SRRF and PySTACHIO (21) are examples of popular single-molecule localisation microscopy software.

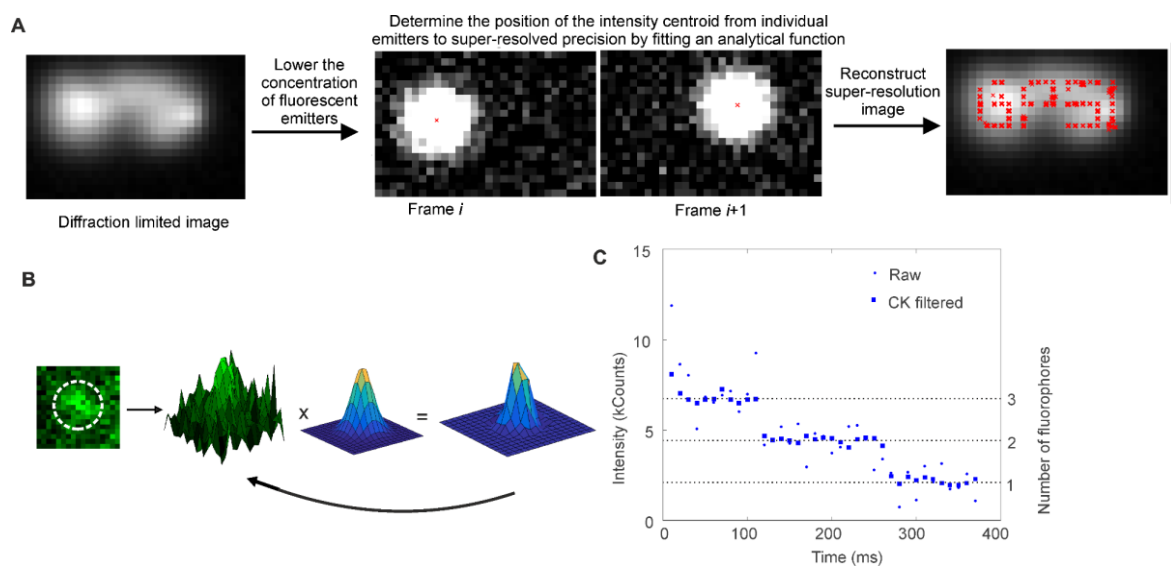

*Figure 4: Principle of single-molecule localisation microscopy utilising dye blinking/switching photophysics. **A**. The panel of the left shows simulated data comprising a frame average over several image frames. The middle two panels show how the dye properties are controlled to ensure only a subset of single molecules emit per image frame such that the mean nearest neighbour separation is much greater than the optical resolution limit, which can be localised to sub pixel precision. The right panel shows how the pinpointing the position pof several such dye molecules over many image frames a super-resolve reconstructed image can be generated, here showing the letters BPSI, standing for the Biological Physical Sciences Interdisciplinary network at the University of York. **B**. Iterative Gaussian masking to determine PSF intensity centroid in a fluorescence image **C**. Step-like photobleach trace of multiply labelled molecule showing raw and Chung-Kennedy (CK) filtered data, a common filtering algorithm which*

> *preserves the edges of step events in noisy data, and so often employed in many different types of single-molecule analysis (22,23).*

2.1.3 Emission-depletion microscopies

In stimulated emission depletion (STED) microscopy the fluorescent dye molecules are excited with one laser and de-excited using stimulated emission via a donut shaped depletion beam in the sample using a wavelength longer than their fluorescence emission. This reduces the area in which fluorescence occurs so that any detected photons must have been generated in the central donut hole. Scanning the beams across the sample enables single molecule localisation down to a transformative molecular precision, with the ultimate limiting being sample photodamage.

Reversible saturable optical linear fluorescence transitions (RESOLFT) microscopy uses a longer-lived conformational state change to de-excite molecules beyond of donut hole, requiring lower laser intensity and so lower photodamage and more rapid sampling through beam multiplexing across a sample. An array of more than 100,000 laser beams has been demonstrated, allowing areas in excess of 100μm x 100μm in living cells to be scanned in under a second.

Minimal fluorescence photon fluxes (MINFLUX) microscopy uses a donut beam for excitation, but no de-excitation beam is required. Fluorescent dye molecules at the beam centre are not excited, and the position where intensity is lowest is the most likely position of the particle. By repeating the measurements over at least three positions form an equilateral triangle around the found position, and by comparing the intensities using weighted trigonometry, sub-nm localisation can be achieved at lower photon numbers than are required for intensity maxima fitting. In principle, the spatial resolution can be arbitrarily improved by increasing the excitation laser intensity provided the dye molecule fluorescence emission is not in a saturation regime, with the only limit then being photodamage to the sample. However, fluorescent dye molecules must still be separated for this method to work.

*2.1.4 3D localization microscopy*

Localisation microscopy has been extended into 3D by fitting the fluorescence emission point spread function (PSF) axially (i.e. orthogonal to the microscope focal plane in the sample) as well as laterally. For a standard optical microscope, the axial PSF width is typically 2-3 times greater than the lateral PSF width of ~half a wavelength and so is relatively insensitive to height differences above and below the focal plane over the typical depth of field of a few hundred nm for a high magnification microscope. However, these small differences can still be used for 3D tracking by focusing on one surface of the sample or using biplane microscopy which utilises a second objective lens, opposite the first, focused on a second plane to sample the PSF of a single molecule at two different axial positions to determine the height.

A PSF itself can also be engineered to better encode 3D information, the simplest method utilising astigmatism microscopy which uses a cylindrical lens in the imaging path to offset the image formed on the camera detector along one lateral axis creating an elliptical distortion of the PSF dependent on the axial coordinate. Combined with STORM, this approach has achieved 30 nm and 50 nm lateral and axial resolution respectively. Use of a phase mask or spatial light modulator in the imaging path, manipulates the PSF to better enable 3D imaging. A common engineered-PSF method uses a double helix shaped PSF comprising two lobes in the sample plane which therefore rotate around each other as the axial position is changed, achieving the best axial localisation resolution for 3D light microscopy imaging techniques. A wide range of PSFs have recently been developed using different axial operating ranges and resolutions including corkscrew, self-bending, saddle-point and tetrapod (Fig. 5).

An emerging method to measure axial distances involves radiative decay. Here, metal or graphene surfaces can perturb the radiative decay rate from fluorescent dye molecules. This process has a sensitive distance dependence and so can be used as quantitative metric to determine distance of the molecule from the surface.

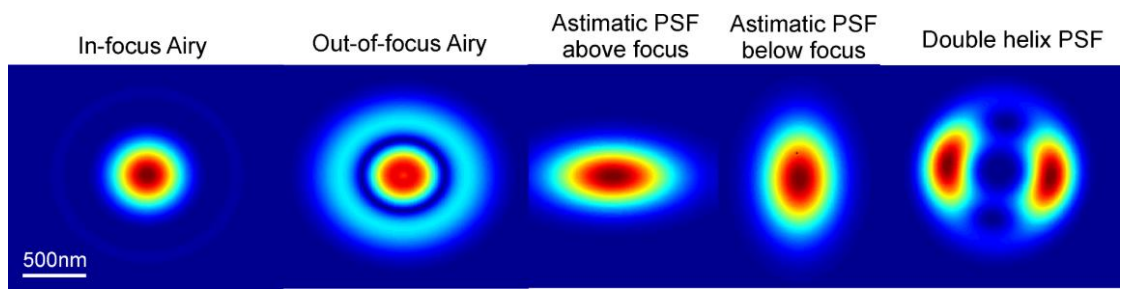

*Figure 5: Intensity heatmaps showing simulations for a range of different PSF geometries relevant to 3D localisation microscopy.*

*2.1.5 Quantifying single-molecule interactions using Förster energy resonance transfer*

Förster energy resonance transfer (FRET) results in non-radiative transfer of energy between two nearby molecules via an electronic resonance effect of their mutual outer molecular orbitals. The length scale of this effect is over ca. 10 nm and so is a good signature for putative interaction between the molecules, which are commonly though not exclusively fluorescent and used as single-molecule tags on biomolecules under investigation. The efficiency of energy transfer is greatest if the excitation and emission spectra of two such fluorescent molecules overlap significantly, such that the shorter wavelength emitter acts as a donor to excite the longer wavelength emitter denoted the acceptor. The efficiency of energy transfer is inversely proportional to the sixth power of the distance between the corresponding electric dipoles of the two molecules, enabling very sensitive distance measurements over approximately 0-10 nm. Using alternating laser excitation (ALEX) of the donor and acceptor excitation wavelengths facilitates robust quantification of both the number of molecules engaging in FRET and the FRET efficiency without the need for heuristic correction factors to account the effects of crosstalk and bleed-through between donor and acceptor detector channels. The first observation of single-molecule level (smFRET) used scanning near-field optical microscopy of a short DNA test construct to separate the FRET dye pairs. It is more common now to apply smFRET using farfield approaches, such as in diffusing molecular species inside a confocal excitation volume. This approach is similar to fluorescence correlation spectroscopy (FCS) which, although not a de facto single-molecule tool, can generate estimates for the effective diffusion coefficient of fluorescently labelled biomolecules as they diffuse through the confocal volume from autocorrelation analysis of the time dependence of intensity peaks measured as each molecule is detected.

Confocal excitation smFRET has been used to observe initial transcription by RNA polymerase on a DNA molecular track through a "scrunching" mechanism involving stored elastic energy in the DNA. Such an approach generates information of FRET interactions through a relatively short time window of milliseconds or less during which two just interacting molecules diffusing the ~micron length scale of the confocal volume. To access information from longer time scale processes in biology, the fluorescent labelled molecules can be immobilised onto a microscope coverslip surface and excited using TIRF microscopy which gives typical access to time scales which are 2-3 orders of magnitude higher. This was used to detect mRNA as it exits RNA polymerase II in real time. The approach has now been extended to use up to four fluorescent dyes; with two dyes undergoing FRET with have information just on the spatial displacement between the tow, not of its directionality, but with multiple dyes it is possible in principle to triangulate different FRET signals from the same two molecules simultaneous the thus extract dynamic 3D structural data at a single-molecule level. MD simulations have also been valuable for predicting the likely molecular conformational dynamics from smFRET data, extending a FRET toolkit to provide time-resolved structural information in a way that traditional structural biology approaches such as X-ray crystallography fail to do.

*2.1.6 Single-molecule methods to study molecular turnover*

Optical microscopy methods of fluorescence recovery after photobleaching (FRAP) and fluorescence loss in photobleaching (FLIP) are increasingly used to quantify molecular in live cell experiments at a single-molecule precise level. In FRAP, a focused laser is used to photobleach of an area of interest in the sample, typically a live cell, with the subsequent recovery of fluorescence into the bleached area a metric for the diffusion and turnover of fluorescent molecules from the unbleached regions of the sample. In FLIP, an area is continuously bleached and the associated loss of fluorescence from a different area, due to turnover of the bleached molecules, is measured. FLIP and FRAP provide complementary information and can determine molecular diffusion coefficients and the relative abundance of mobile and immobile components of the molecular population. Most measurements from FRAP and FLIP are technically ensemble in relating to summed fluorescence emissions from multiple molecules; however, they can still be used to infer molecular precise stoichiometries and densities. A recent variant method is called single-point single-molecule FRAP in which molecular turnover is determined from a photobleached area of nuclear membrane as transmembrane proteins returning to the bleached region on an individual basis, to build up a measure of the concentration ratio between the inner and outer nuclear membranes.

*2.1.7 Structured illumination microscopy (SIM)*

SIM is another popular super-resolution optical microscopy tool which can enable single-molecule detection sensitivity. It applies patterned light onto the sample to generate interference Moiré fringes that encode higher spatial frequency super-resolution information despite the raw image data itself being diffraction limited. By recording patterned images with typically three different fringe geometries oriented at 120º to each other in the 2D sample plane, one can sum the images in reciprocal space, and inverse Fourier transform to produce a final image with approximately twice the effective spatial resolution than conventional diffraction-limited microscopy, with similar approaches allowing application to 3D information. Non-linear illumination can utilise harmonics in the fringe pattern to extend the spatial precision even further, for example saturated structured illumination microscopy (SSIM) with pulsed lasers gives a resolution of around 50 nm. Non-linear SIM can also be achieved using photoswitchable proteins, requiring lower laser powers than SSIM. The method of Instant SIM (iSIM) develops this technique to 3D, which benefits from improvements in sampling speed via two microlens arrays, a pinhole array and a galvanometric mirror, allowing full 3D imaging at 100 Hz at the expense of lateral and axial spatial precisions of ca. 150 nm and 350 nm respectively. Extending iSIM to incorporate 2-photon excitation increases the penetration depth and makes the method suitable for thick specimens such as tissue sections. Several recent developments correlative SIM with other techniques, such as TIRF.

*2.2.8 Electron microscopy advances*

Recent developments in cryo-EM are proving enormously fruitful in enabling imaging of the structures of relatively complex single biomolecules and molecular complexes which are too challenging to crystallise and/or which exist in dynamic and heterogenous states. Such complexes can be relatively large assemblies, and cryo-EM often the capability to determine structures in their native state, including many membrane proteins which have proved too challenging using NMR and X-ray crystallography. Developments with reliable pattern recognition tools and data compression algorithms developed from AI have helped significantly with dealing with one of the key challenges of the generation of significant volumes of data to analyse. Operating below the diffraction limit of optical microscopes, EM in general can resolve details at the level of Ångströms and is therefore an unparalleled tool for structure determination. But, as with X-ray crystallography, a snapshot of a biomolecule on its own cannot give the full picture. One method for probing dynamics directly using EM is ultrafast electron microscopy (4D UEM) for which it has been possible to image in femtosecond resolution protein vesicles combined with nanometre precision spatially, and to image a full *Escherichia coli* cell.

Standard EM approaches do not traditionally need to tag biomolecules of interest through genetic means or otherwise, although immuno-EM methods do achieve this with specific antibodies which have gold particles of typically a few nm in diameter conjugated as an electron dense contrast reagent. Correlated imaging involving EM and fluorescence, discussed later in this review, have some value in combining the high spatial precision of EM with the specificity of labelling a range of different cellular structures with fluorescence. Demonstration of a genetic tag which enabled this was published in 2011 in which the tag generated singlet oxygen which catalysed a reaction that was then visible in an EM image. This tag was shown to colocalise with simultaneously obtained fluorescence data, giving structural and localisation information previously unobtainable.

## 2.2 Single-molecule electrical conductance

Electrical conductance measurements have been used to study cell membrane channel proteins previously; however, this approach is increasingly now applied to more diverse questions, in particular sequencing of DNA and, potentially, working towards direct sequencing of proteins.

### 2.2.1 Electrical patch clamping

The patch clamp has been a workhorse of experimental cell physiology since the 1970s, typically comprising a narrow glass pipette electrically sealed onto a small area of cell membrane containing only one ion channel, and the flow of ion currents through the opening can be recorded. Single channel patch clamp has been used to study ion channel proteins involved in electrical conduction in nerves and neuromuscular junctions. More recent applications include biological processes such as those in insect olfactory receptors, those of direct relevance to impaired muscle contraction as occurs during atrial fibrillation, and the mechanisms of anion conduction by coupled glutamate receptors. Importantly, the patch clamp is also finding new applications in correlative techniques, as discussed later in this review.

### 2.2.2 Measuring molecular signatures through nanopore ion fluxes

Pores of comparable diameter to single molecules, so-called nanopores, can enable single-molecule detection via the extent to which a molecule will occlude the nanopore aperture as it passes through it, thus resulting in a transient but measurable drop in electrical current. The molecule of interest is introduced at one side of the pore and a voltage applied, forcing molecules, which in general have a net non-zero electrical charge, through the pore via electrophoresis. As each molecule translocates through the pore and occludes its aperture characteristic drops in electrical current are measured dependent on size and shape of the molecule. The pores themselves can be solid state, such as silicon dioxide, graphene and DNA origami. The most popular pore material currently is composed of protein; a commonly used natural protein nanopore is that of α-haemolysin which is a toxin produced by *S. aureus* designed to punch holes in the cell membranes of competing surrounding cells. These protein nanopores are now regularly used in DNA sequencing, since different nucleotide bases of DNA will produce slightly different drops in electrical current as the individual bases pass through the nanopore while the whole molecule translocates like thread through the eye of a needle through the nanopore. The company Oxford Nanopore have developed a USB-sized chip-based nanopore sequencer with long read lengths which can produce several GB data/day. Nanopores have also been used to detect mRNA molecules from lung cancer patients and to detect proteins directly and protein interactions with nucleic acids; more recently, the unfolding kinetics of single proteins has been measured.

*2.2.3 Scanning ion conductance microscopy*

Scanning ion conductance microscopy (SICM) is a scanning probe technique which uses ion flux to measure the topography of a sample due to an increase in electrical resistance as the surface is approached. The probe is a nanopipette of diameter 10-30 nm. As the tip is moved to within its own diameter from the biological sample being scanned then the ion flow is impeded. With fast feedback electronics to those used for atomic force microscopy (AFM), discussed as a technique later in this review, this level of electrical resistance can be used to maintain a constant distance between tip and sample and so generate topographical data as the tip is laterally scanned over its surface with the spatial resolution limited by the diameter of the nanopipette. This is worse than AFM but SICM causes less mechanical damage to the sample. The technique has also been adapted to be combined with single-molecule folding studies of fluorescent proteins: here the same nanopipette is used primarily to deliver a chemical denaturant to unfold, and therefore photobleach, a single fluorescent protein molecule, prior to their refolding and gaining photoactivity, and so this method can be used to study the kinetics of these processes.

## 3. Measuring and controlled molecular force and torque

All cellular processes ultimately involve the application of forces between single molecules, and/or corresponding torques. As already discussed, a suite of different molecular motors are involved, such as myosin pulling against actin to result in muscle contraction; kinesin translocating along microtubule tracks to transport cellular cargos;

during DNA replication, DNA supercoils build up due to the translocation of DNA polymerase, which needs to be relaxed by a class of enzyme called a topoisomerase to a less stressed conformation for continued replication to then proceed; ATP synthase creates ATP from adenosine diphosphate (ADP) by mechanical rotation, as well as monitoring DNA topology, phase transitions and even knots at a single-molecule level.

*3.1 Methods using light to manipulate single molecules*

Photons possess both linear and angular momenta. When they hit an object, momentum transfers to the object because due to refraction, reflection and absorption processes, resulting in a net force and corresponding torque applied on the object, which underpins the essential physics of optical tweezers.

*3.1.1 Optical tweezers*

Optical tweezers (OT) use a high numerical aperture (NA) objective lens to focus a collimated laser beam in the vicinity of a refractile object such as a ~micron diameter glass or plastic bead, resulting in a potential energy minimum close to the laser focus, though in practice marginally in front due to forward radiation pressure. Fig. 6A shows a basic laser tweezers setup. The bead is optically trapped by the laser so by moving its focus one can manipulate the bead and thus any biomolecules which are conjugated to the surface of the bead. NIR lasers act as the typical laser source which, even with high powers used up to ~10 W cause relatively minimal photodamage to biological molecules compared to visible wavelengths lasers of comparable power. Simple OT use a Gaussian beam profile so that the intensity maximum is at the centre of its waist. Force transduction can be understood using simple ray optics. The bead, having higher refractive index than the surrounding liquid, will experience gradient forces from the peripheral part of the beam which push the bead towards the laser focus where light intensity is at a maximum. The radiation pressure from the central part of the beam pushes the bead forward slightly downstream of the focal waist. Scattering and the gradient force contributions to trapping forces mostly emerge from net refraction and its associated change in momentum of the altered photon path direction effects rather than reflection or absorption. The maximum optical force that an optical trap can apply to along any of the three spatial directions is:

$$F = Q\frac{nP}{c} \qquad (5)$$

Here $Q$ is a constant between 0 and 1, $n$ the refractive index of the surrounding liquid; $P$ the laser power of the laser incident on the bead, and $c$ is the speed of light in vacuum. Four example scenarios of different bead displacement relative to the trap centre are depicted in Fig. 6B-E. Here, the same three rays are shown to illustrate the gradient and scattering forces. In Fig. 6B the bead is at equilibrium such that the net force is zero. Fig. 6C shows the bead at the centre of the focal plane with zero net gradient force so the only force from the laser is the upward scattering force so the bead is pushed upward. Fig. 6D shows the bead below the focal plane with scattering and gradient forces pointing upward. Fig. 6E shows the bead positioned laterally off the optical axis, and here the net gradient force pulls the bead back towards the left towards the trap centre despite the scattering force pushing the bead to the right since it is weaker than the gradient force. For a Gaussian laser beam intensity profile, the scattering force partially balances out axial gradient forces, so $Q_{axial} < Q_{radial}$. Thus $Q_{axial}$ limits $Q$ of the optical trap. In an electric dipole approximation description, instead of the more qualitative ray-optics description, the bead is treated as a dielectric particle which is electrically polarised by the optical field in the laser trap. This optically induced electric dipole interacts with the electromagnetic field of the light to move along the field gradient, resulting in the trap applying a restoring force to the bead in all three dimensions.

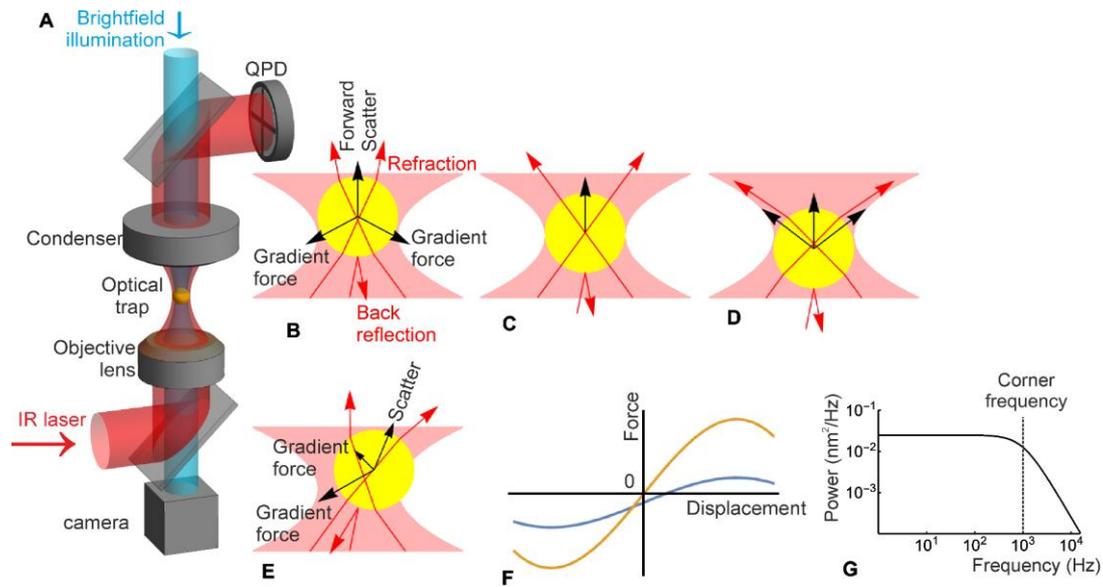

*Figure 6: Carton illustrating OT and associated forces. **A.** Stripped down depiction of the typical optical components in an OT; a collimated near IR laser beam enters the objective lens back aperture and is focused to a diffraction-limited confocal volume (manifest as a focal waist in the 2D sample plane), before being re-collimated by the condenser. The emergent IR beam is imaged on a quadrant photodiode (QPD). **B.-E.** show forces applied on the bead: **B.** in force equilibrium, **C.** at the centre of the laser focus, **D.** below the centre and **E.** displaced to the right of the optical axis. F. Shows two force-displacement curves for vertical (blue) and horizontal (orange) displacements. These are linear near equilibrium positions, suggesting a Hookean-spring regime, its gradient being the trap stiffness. **G.** is the theoretical Lorentzian function for a power spectrum of a trapped bead; the corner frequency is approximately at the 'corner' of the log-log plot and is labelled with a dashed line and is directly proportional to trap stiffness.*

Trapped bead positions are measured via laser interferometry either from the trapping light or a separate tracking laser source, imaging the back focal plane of the condenser onto a quadrant photodiode (QPD), which can be calibrated to field a force response

(Fig. 6F). The trapped bead in a thermal reservoir of surrounding water molecules has its movement described by the Langevin equation:

$$m\ddot{x} + \beta\dot{x} + kx = F_{\text{thermal}} \tag{6}$$

Here $m$ is the bead mass of the bead, $\beta$ the hydrodynamic frictional drag on the bead, $k$ the optical trap stiffness and $F_{\text{thermal}}$ the Langevin force due to stochastic Brownian motion. Upon Fourier transformation, this equation turns into a Lorentzian shaped curve in the frequency domain:

$$S(f) = \frac{k_\text{B}T}{\pi^2\beta(f^2 + f_0^2)} \tag{7}$$

Here $f_o$ is the corner frequency (Fig. 6G), a point at which the power spectral density drops to 50% of that at zero frequency and it is related to the stiffness of the trap by $k = 2\pi\beta f_o$ and so can be used to obtain $k$. A QPD is used instead of a camera detector due to the much higher bandwidth required to generate a well sampled power spectrum over a broad frequency range. The trap stiffness can also be estimated from the gradient of the force-displacement response (Fig. 6F).

Laser beam profiles can be generated to use higher modes beyond a simple Gaussian, so-called Laguerre-Gaussian (LG) mode beams, which can improve the effective trapping stiffness but are technically less easy to implement. Similarly, Bessel beams can be used which have an intensity profile defined by a Bessel function of the first kind:

$$I(r) = I_0 J_n(k_r r) \tag{10}$$

Again, technically more difficult to implement, Bessel beams have some advantages in being non-diffractive and self-reconstruction after being partially occluded with an obstruction, so have potential applications deeper into biological samples.

*3.1.2 Controlling torque using light*

Transfer of angular momentum can be used to rotate optically trapped objects. The simplest approach to rotation with light is trapping two points on an object with two separate optical traps and rotating the traps around each other. More commonly Bessel, Laguerre-Gaussian or other beams that contain intrinsic orbital angular momentum are used to impart angular momentum on probe particles leading to their rotation at a constant torque. Alternatively, linearly polarised light can rotate a birefringent probe by using a defined angular displacement.

*3.1.3 Holographic OT and lab-on a chip OT*

Holographic optical tweezers (HOT) use of holographic light modulation to create a single beam intensity profile that can trap up to several hundred microbeads in 3D. The required light modulation can be generated via nanofabricated diffractive optical elements or spatial light modulators (SLMs). Soon after its inception, HOT were capable of manipulating hundreds of micro- dynamically. Compared to traditional tweezers, HOT provides has increased flexibility and adaptability, but currently there is not direct way to determine forces on each separate trap directly.

A lab-on-a-chip (LOC) devices can be used to generate optical trapping fields through microfabrication of chip surfaces. These can create dielectric nanostructures to shape light and generate optical trapping potentials for single micron scale refractile particles. This can be used for high-throughput screening with the added benefit of having associated microfluidics to control sample fluid environments.

*3.2 Magnetic force techniques*

Magnetic forces can manipulate micron sized magnetic particles for use in measuring and controlling force and torque in single-molecule biophysics, depending on the type of magnetism used. With ferromagnetism or super-paramagnetism the magnetic forces can are generated from the interaction between a magnetic field vector, **B**, and the magnetic dipole moment vector **m** of a suitable magnetic particle, such that the magnetic force vector $\boldsymbol{f}$ is proportional to the $\boldsymbol{B}$ field gradient, whereas the torque vector $\boldsymbol{\tau}$ aligns with the field itself:

$$\boldsymbol{f} = \nabla(\boldsymbol{m} \cdot \boldsymbol{B}) \tag{11}$$

$$\boldsymbol{\tau} = \boldsymbol{B} \times \boldsymbol{m} \tag{12}$$

*3.2.1 Magnetic tweezers (MT)*

The equilibrium position for a magnetic particle, typically a micron scale bead, in MT is the local gradient maximum, which is inside either a permanent magnet or electromagnet and so the magnetic particle never reaches force equilibrium. The magnetic forces do not 'trap' the bead but fix it in position by feedback loops that constantly adjust the field or by balancing the forces from a tethered biomolecule, such as DNA either using a permanent magnet (Fig. 7A) or electromagnet (Fig. 7B) which can change the level of exerted molecular force by changing the height of the coverslip relative to the magnet (Fig. 7C), as well as changing the torque through rotation of the magnetic B-field. Magnetic forces can easily reach biologically relevant values of tens of pN, and the ability to apply external torque to a magnetic bead is the key strength of MT (Fig. 7D), which is easier to technically implement than optical tweezers That being said, there are still challenges in providing very high magnetic field gradients for very large forces to be applied in magnetic tweezers.

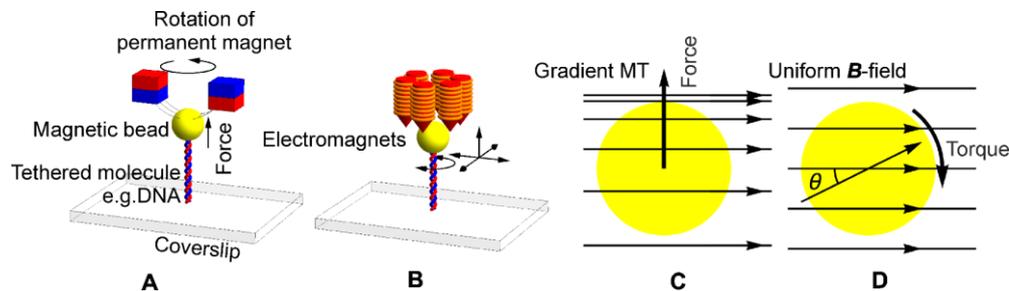

*Figure 7: Schematic of MT configurations for force and torque generation. **A.** Permanent magnet MT, shown here with DNA is tethered to a microscope coverslip. **B.** Electromagnets can also be used to rotate the magnetic field and applying a gradient force axially. **C.** Magnetic force in a gradient trap points towards regions of higher field density. **D.** In a uniform **B**-field the gradient, and so the force, is zero, but rotation of the B-field can generate torque as the bead's magnetic moment aligns.*

*3.2.2 3D control of MT*

The MT shown in Fig. 7 apply force in only the positive axial direction and torque around this axis. Magnets are held by a motorised arm that can move the magnets closer to or further away from the sample along the vertical axis. MT of other translational and rotational degrees of freedom have been designed to meet the needs of biological molecule manipulation, e.g. MT that illustrated in Fig. 7B has poles at six corners above the bead so can apply radial force in addition to upward pulling. Theoretically four poles can apply radial force as well; Huang et al. designed such an MT, but they had a matching four poles underneath the sample so full 3D translation and rotation was possible. Spatial constraints are a major design consideration for MT: a typical commercially available light microscope which uses a high numerical aperture objective lens for high resolution imaging, as is most relevant to single-molecule biophysics, has limited space available between the objective lens and the sample, and sample and the condenser lens, not ideal, as they stand, for implementing larger MT designs. The setup of Fig. 7B can be incorporated into a commercial microscope, Huang's design needs to be incorporated into a bespoke microscope. Miniature MT use electromagnetic tweezers small enough to fit into the flow cell.

Helmholtz electromagnetic coils are the simplest geometry to achieve a near-uniform magnetic field. Two pairs can be arranged perpendicular to each other can rotate the bead along one axis whereas three pairs can rotate along all three directions. Although one pair cannot rotate the bead by itself, it has found applications in oscillating the beads. Also, ultimately the impedance of electromagnetic coils limits dynamic measurements with this type of magnetic tweezers. Loenhout et al. used such an approach to supercoil a DNA molecule and extend it in the transverse plane for fluorescence imaging, while Shepherd et al. (24) combined this type of strategy with optical tweezers to allow complete ability to extend, twist and image single tethered molecules such as DNA, denoted Cominatorial Optical

*3.2.3 Measuring torque*

To measure torque in MT requires estimation of the angle between the bead's magnetic moment and the **B**-field. Traditional MT have issues associated with large torsional stiffness since the **B**-field gradients capable of applying physiologically relevant forces exert typical torques at easy to measure angular displacement that can be several orders of magnitude above biologically relevant values. Such high-stiffness MT cannot be easily calibrated for torsional stiffness at physiological relevant values due to the challenges of detecting relevant small angular displacement in the bead that are required for physiological levels of torque. Methods to overcome this problem have included using a transversely orientated rod adhered to a bead to amplify the angular displacement, as well as use of small fluorescent beads as fiducial markers on the bead surface (24).

*3.2.4 Nanofabricated, high throughput and hybrid MT*

Traditional MT instrumentation has a length scale of a few cm and are often built around home-made microscopes rather than being commercially available. However, recent progress has been made to lithographically embed MT in a flow cell such that it integrates with any existing microscope design. Fig. 8 illustrates an example of nanofabricated MT with six poles in the same plane (yellow) and a ring of coil (or 'ring trapper') lying flat in the centre (red), reported by Chiou et al. The poles apply forces in the 2D plane of the magnets while the ring provides an axial attractive force. A trapped superparamagnetic or ferromagnetic bead can then be rotated. The pole to sample distance can be reduced to a few hundred microns which is an order of magnitude smaller than traditional MT and since the $\boldsymbol{B}$ field scales with $1/r^2$ this requires smaller currents and a reduction in technical issues associated with localised heating. Also, the increase in local B-field magnitude implies that smaller magnetic beads can be used for generating equivalent torques which therefore have smaller associated frictional drag and thus enable more rapid torque experiments to be performed which may be important for exploring some fast biomolecular motor conformational changes. Similar designs reported by Fisher et al. shown in Fig. 8B use alternately raised poles to enable translational and rotational control along all spatial directions.

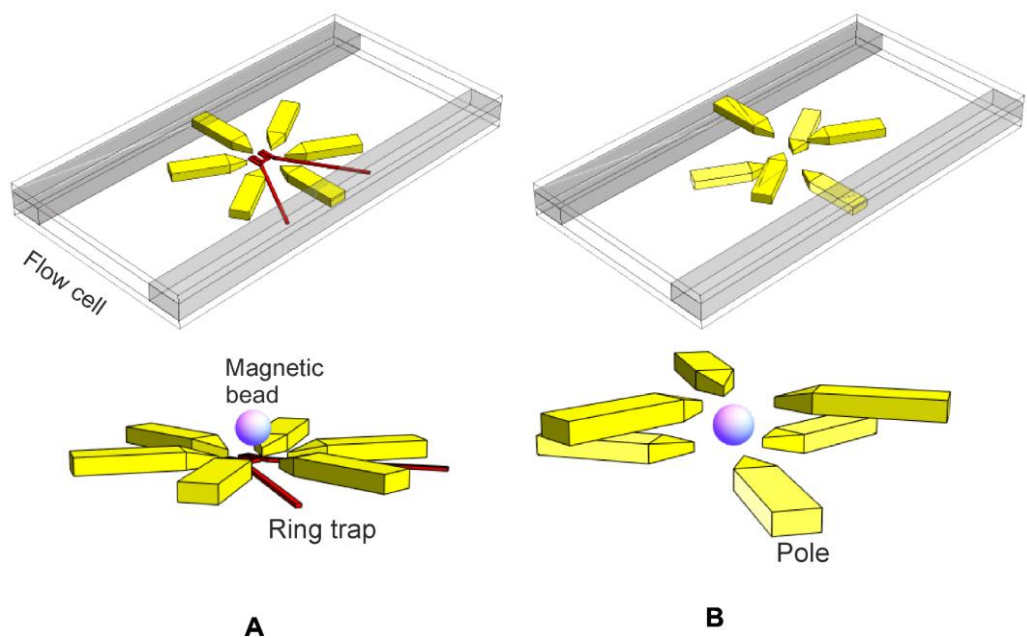

*Figure 8: Nanofabricated MT designs. **A**. Six poles in the same plane, each independently controlled coil with a pole piece in the centre. There is also a central wire circle to provide vertical force. **B**. MT with three poles that are in one height and three raised relative to enable 3D translational and rotational control of the magnetic bead.*

Traditional MT have an advantage over nanofabricated MT in being intrinsically high throughput since the **B** field strength is uniform over a relatively large volume in these devices, typically several mm, and so presents opportunities to perform experiments on several magnetic beads simultaneously (Fig. 9). More recently, the transformative COMBI-Tweez technology already alluded to in this review has been developed utilising MT, OT and fluorescence imaging in one instrument to enable independent twisting and extension of chiral biopolymer molecules, tested on DNA as one such model molecule of this class, correlated to visualisation of their topology in real time (24).

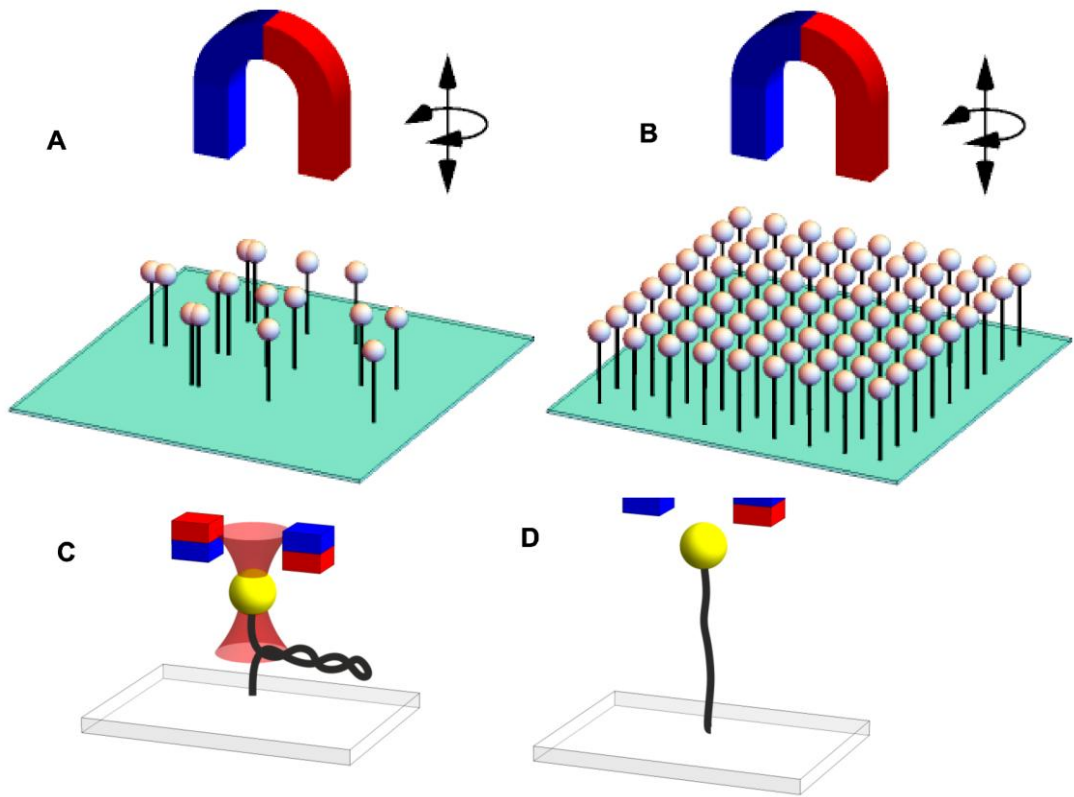

*Figure 9: Schematics of multiplexed and hybrid assays using magnetic tweezers. **A.** Here, DNA-tethered beads randomly diffuse and tether to the microscope coverslip surface. **B.** Alternatively, spatial micro-printing of conjugation chemicals onto the coverslip surface can enable spatially periodic placement of DNA-bead tether complexes which, compared to random spontaneous tethering, which has an order of magnitude higher yield of usable tethers in the same field of view. **C.** and **D.** showing hybrid combinations of MT and OT, such that in **C.** the OT applies a downward force opposing the gradient force from the MT resulting in a DNA plectoneme forming, and **D.** when the OT is subsequently turned off, the plectoneme unravels.*

*3.2.5 Using acoustically generated forces to manipulate single molecules*

The air pressure gradients created by acoustic waves can be used to trap microscopic particles inside the anti-nodes generated from the associated standing waves, such as surface-functionalised beads which can then be used for single-molecule tethering experiments in a similar manner to OT and MT The wavelength of these standing waves can be adjusted to control the movement of these trapped beads. Shi et al. focussed multiple microbeads in a narrow line via acoustic waves, placing two pairs of interdigital transducers at right angles traps the particles in 3D. Here, the beads were trapped vertically by balancing gravitation, buoyancy and viscous drag forces. The associated instrumentation – a setup named acoustic tweezers (AT). The energy density is ~10 million times lower than that of optical tweezers resulting in a marginal advantage in not raising the temperature as much (with OT, a typical temperature increase for a relatively stiff trap is ~2 K), but the innate high throughput capabilities due to the multiple nodes and antinodes present in the chamber is a more significant advantage. AT have even been applied to the selective control of microbeads *in vitro* as well as *in vivo*. Developments are helping to optimise acoustic holography, which in the future may conceivably enable 3D trapping capability.

**3.3 Surface probe microscopy (SPM)**

There are at least ~20 different surface probe microscopy (SPM) relevant to single-molecule biophysics. These all use a physical readout of force interaction between a probe and a surface to map out the molecular surface topography. The most useful common and useful SPM technique to single-molecule biophysics is atomic force microscopy (AFM).

*3.3.1 Atomic force microscopy (AFM)*

In AFM, a tip is attached to a thin, compliant microscopic metal cantilever to probe the surface of a sample. A laser beam is directed onto the back of the cantilever which reflects the beam to a photodiode to measure the beam's deflection and hence is a measure of the cantilever bend which, after appropriate calibration, can be converted to an equivalent force. The AFM tip is of microscopic length scale but with a nanoscale radius of curvature normally silicon based such as silicon nitride and interacts with the sample primarily through van der Waals attraction and Coulomb repulsion resulting in cantilever bending depending on the distance of the sample from the tip. AFM has an Ångström level spatial resolution axially, laterally typically an order of magnitude higher limited by the radius of curvature of the tip and the distortion of the relatively soft biological sample due to the forces imposed by the tip as it is scanned laterally across the sample.

AFM can be applied to imaging and force measurements. For imaging, the cantilever performs a lateral scan over the sample surface and generates a reconstruction of the topography. Most implementations of AFM have relatively low temporal resolution corresponding to at least several seconds per reconstructed image frame due to the mechanical nature of the imaging mode. However, high-speed AFM (HS-AFM) has been demonstrated with less than 100 ms per image frame, allow for example the stepping translation of single myosin motors on actin tracks to be visualised. In force spectroscopy mode, the AFM tip moves vertically via a piezo electric actuator. The cantilever is tethered to the biological molecule so it can then be pulled/pushed to determine a force-displacement curve, which can reveal conformational changes of the biological sample, such as the unfolding dynamics of domains within a single filamentous molecule.

AFM has been used in the detection and localization of molecular recognition events, cell surface probing to single-molecule resolution, and studying the structural properties of proteins and nucleic acids. Recent developments in AFM include externally tuning the oscillator's response characteristics for robust position control and functionalisation of the cantilever tip has also been modified for higher resolution.

*3.3.2 Dielectric spectroscopy*

A variant of AFM is dielectric spectroscopy that measures the local electric dipole moment (i.e. the electrical permittivity) of a sample in response to an applied electric field, varying over a range of input frequencies. The technique can measure molecular fluctuation directly and it has high enough sensitivity to precision measure thermal expansion. Biological applications include measuring molecular interactions, detecting cancerous cells and more generic cells including bacteria and yeast. Label-free microfluidic biosensors have also been contrived to apply dielectric spectroscopy.

*3.3.3 Electrostatic force microscopy (EFM)*

EFM uses core physics concepts similar to that of AFM, though here it is electrostatic forces that are specifically measured. A cantilever with an electrically conductive tip is used to measure the electrostatic force from a biological surface. A voltage is applied between the tip and the sample so the local charge distribution over the surface causes the tip to experience either attractive or repulsive forces, which bends the cantilever towards or away from the surface. The extent of cantilever displacement is read out in the same way to that of AFM. In non-contact mode, the cantilever is placed at a distance far enough away from the sample such that the net force is always attractive. The changing electric force varies the resonance frequency of the cantilever, which is converted to electro-topography and electrical charge density measurements. Conversely, in contact mode, the tip-sample distance is kept constant; the mean forces exerted on the sample are higher resulting greater spatial resolution at the risk of more sample distortion. EFM has been applied to biological samples for imaging photosynthetic proteins, visualising charge propagation along individual proteins, synthetic biological protein nanowires, electron transfer between microorganisms and minerals and cell interactions within microbial communities.

*3.3.4 AFM "cut-and-paste"*

AFM cut-and-paste can be used for the assembly of biomolecules at surfaces by using AFM to manipulate and reposition individual molecules. This approach typically combines AFM with DNA hybridization technology to pick up individual molecules from a depot area on the sample surface and arrange them on a construction site one by one. Anchors and handles usually comprise DNA but alternatively a broad range of ligand receptor systems may be employed. Kufer et al. used complementary DNA strands to manipulate and reposition functionalised DNA oligomers to Ångström level precision. Each molecule transfer can be monitored with fluorescence microscopy combined with force spectroscopy, and this approach can also be used to deposit fluorescently labelled molecules in tagged in pre-defined patterns.

## 3.4 Electrical force manipulation tools

Several biophysics techniques which utilize electrical forces, and many of these are relevant to single-molecule science, in particular methods which can rotate molecules and dielectrophoresis tools which can translationally manipulate them.

### 3.4.1 Electrorotation tools

Electrorotation has similarities to MT as a biophysical single-molecule rotation technique. Here, a microbead with a permanent electric dipole moment is rotated by application of a 3D electric field generated from microelectrodes constructed inside a flow cell. The physical principle of operation is that mobile electrical charges take a finite time to reorientate in aligning to an AC input field and thus for a rapidly oscillating E-field there is a phase lag between the bead's electrical dipole and the input field. This lag results in a torque, thus bead rotation. Rowe et al. (1) applied rotation to bacterial flagellar motors via electrical charged beads and were able to measure rotations rates in excess of 1 kHz.

### 3.4.2 Anti-Brownian ELectrokinetic (ABEL) traps

Brownian motion of single biomolecules in solution poses challenges to their tracking and visualisation; there is typically a ~millisecond maximum sampling window available before molecules diffuse out of a region of interest of a typical high spatial resolution optical microscope. Anti-Brownian ELectrokinetic (ABEL) traps use an electric field to confine the molecules assuming they have a net electrical charge, which is generally the case. Wang and Moerner implemented ABEL to measure the diffusion coefficient and mobility of single trapped fluorescent proteins and oligomers. The current major limitation is that the electrical trapping forces are principally 2D and so eventually a biomolecule will diffuse axially out of the microscope's depth of field, though progress is being made to develop 3D ABEL capability.

## 4. Future opportunities and challenges in single-molecule biophysics

Recent developments in single-molecule biophysics have resulted in a suite of intriguing future opportunities and challenges. These and their wider implications are discussed below.

### *4.1 Correlative single-molecule biophysics approaches*

New physical and biological insight is likely to emerge from combining multiple existing single-molecule techniques in the same investigations; if the acquired data streams from each technique are in effect orthogonal and synchronised then the likelihood for true positive single-molecule detection, as essentially the product of two or probability distributions, is significantly higher compared to any of the single techniques used in isolation. Here we review recent developments in such correlative methods relevant to single-molecule biophysics.

#### 4.1.1 AFM and optical microscopy

AFM and fluorescence imaging have been combined in single instruments for decades, but the advancement to AFM with single-molecule fluorescence detection is more recent needing to overcome technical challenges of limited space around an optical microscope in which to integrate an AFM, which requires a highly stable platform to achieve high resolution. For thin, transparent specimens AFM can be integrated from above the sample with fluorescence imaging from below (Fig. 10A). For thicker specimens there can be uncertainty as to whether the same molecule is being imaged in both modalities. A second challenge is that AFM, as a rule, requires a high density of target molecules, due to the difficulty in directing the AFM tip detecting individual single molecules; this in direct contrast to single-molecule fluorescence microscopy in which fluorophores are ideally separated by more than the diffraction limit to be localised efficiently by tracking algorithms. A third challenge is that stray light used for measuring AFM cantilever deflection can interfere with fluorescence detection, contributing to the background noise.

This integrated approach has been used with TIRF geometry for single-molecule fluorescence microscopy but limited due to the finite depth of field of fluorescence excitation of ~100 nm that can be achieved. Correlative AFM and super-resolution single-molecule fluorescence microscopy on structural elements in live mammalian cells has been shown demonstrated sequentially. AFM-FRET microscopy (Fig. 10A inset) allows mechanical changes to be monitored over time simultaneously, e.g. the force resulting from a conformational change can be measured alongside the corresponding conformational displacement change measure using FRET. An alternative to labelling the protein being manipulated by AFM is to perform an AFM study on an enzyme that produces a fluorescent product and then to monitor fluorescence as a metric of catalytic activity. There are also similar useful combinations involving optical tweezers and fluorescence.

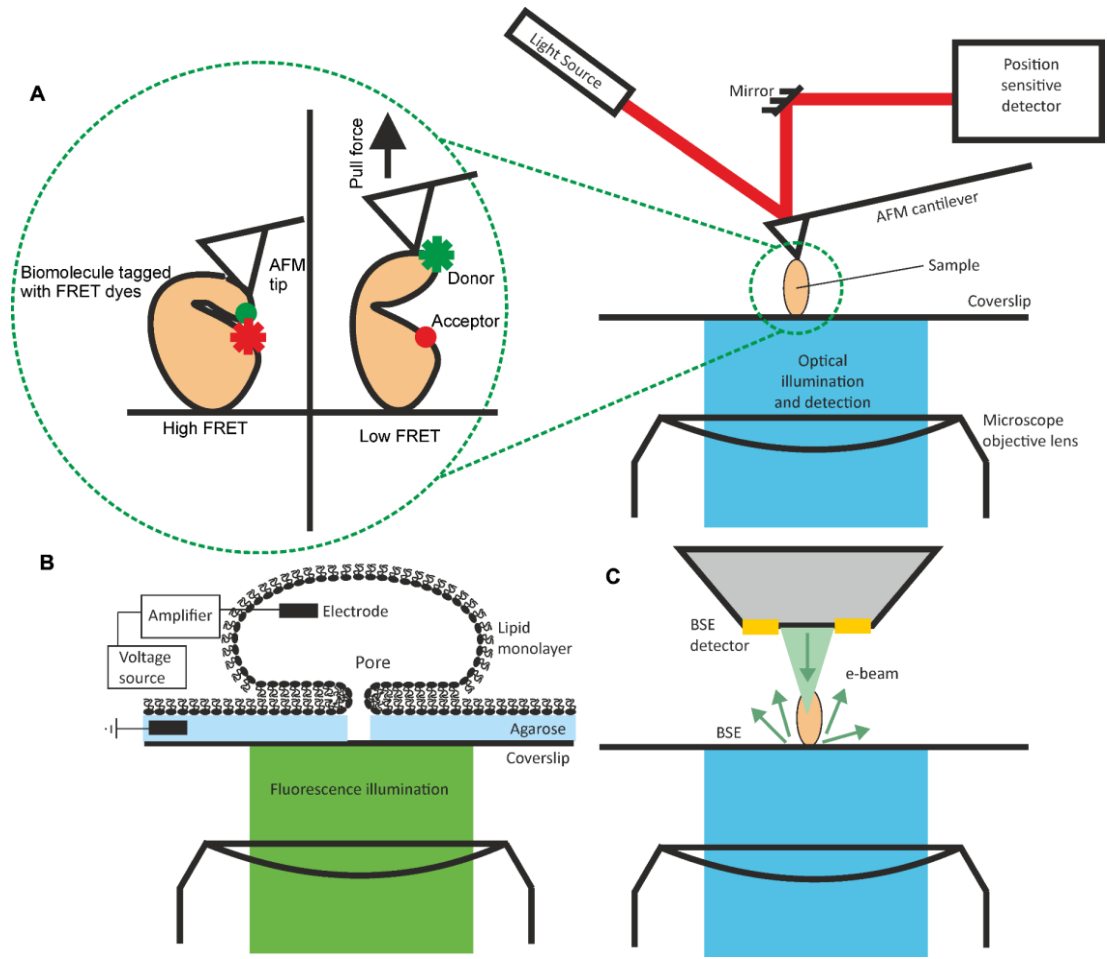

*Figure 10: Correlative imaging techniques. **A.** AFM-FRET microscopy; single-molecule fluorescence microscopy performed in inverted mode, enabling top-down AFM. As target molecule extends, FRET efficiency changes result in increased longer wavelength emission of the acceptor and decreased shorter wavelength emission from the donor dye. **B.** Cartoon of a combined patch clamp and fluorescence imaging using a droplet interface bilayer; droplet contains calcium and calcium sensitive fluorescent dye. The electrodes can measure voltage across the lipid bilayer area, illuminated with the laser which can be switched to TIRF microscopy to increase image contrast. **C**. Cartoon of CLEM; back-scattered electrons (BSE) are detected to perform EM imaging, while fluorescence microscopy is performed in inverted mode.*

*4.1.2 Patch clamp and single-molecule fluorescence imaging*

One of the earliest correlative biophysical techniques was combined fluorescence microscopy and patch clamp electrical measurements (Fig. 10B), performed since the late 1990s, however early experiments struggled due to the mismatch in temporal resolution available with electrical and optical recording. With the increase in speed and sensitivity of electron-multiplying CCD cameras, coupled with other technological developments, the technique is undergoing a renaissance. Patch clamp can be combined with different single-molecule fluorescence techniques such as FRET or TIRF microscopy to study different membrane transport processes. Patch clamp combined with FRET is ideal for studying molecular conformational changes of ion channels following ligand binding, combined with electrophysiology measurements. By placing a FRET donor and acceptor on the channel subunits, high FRET signal occurs when the channel is closed. States with low FRET signal but no ion current flow identify intermediate states between the channel opening and closing.

The early combined patch clamp and fluorescence techniques offered spatial information with single pore resolution, but until recently were mainly applied to calcium channel investigation via fluorescence detection of calcium using TIRF microscopy on large *Xenopus* oocytes (egg cells from a large toad often used as a model organism), due to the requirement of large vesicles to apply a patch clamp and the availability of calcium sensitive fluorescent dyes compared to the lack of sensitive reporters for other physiologically relevant ions such as potassium. Calcium sensitive dyes such as calcium green-1 dextran and fluo-8 AM increase fluorescence emission flux on binding to calcium on a similar time scale to the calcium concentration changes due to gating events, facilitating *in vivo* imaging of neuronal networks. Potassium channels have been studied *in vitro* using the dye Asante Potassium Green 4, although the dye's relatively slow response limits the utility of these measurements; it takes ~seconds for dye to reach maximum signal after a millisecond time scale gating event.

Advances in making stable lipid bilayers and droplet interface bilayers, coupled with improvements in microfabrication are enabling the study of pore formation that does not involve channel proteins; this study used a geometry similar to Fig. 10B but with a microfabricated substrate which creates multiple droplets on one chip, enabling high throughput. Electrical data is correlated with the fluorescence imaging data from each pore visualised.

*4.1.3 Correlative light and electron microscopy (CLEM)*

Although EM has better spatial resolution that optical microscopy it lacks labelling specificity. Correlated light and electron microscopy (CLEM, Fig. 10C) achieves the best of both worlds with the caveat of several challenges. EM samples are typically chemically fixed, stained and embedded in plastic/wax for sectioning. Special procedures are then followed to preserve fluorescence of normal probes. Alternatively, cryo-EM is used to flash-freeze and preserve fluorescence. Samples can be imaged separately on EM and optical microscopes using finder grids including fiducial markers. Recently, several integrated CLEM instruments have become available commercially which either move the sample between optical and electron imaging modes, preserving registration, or contain paraxial optical and electron imaging.

## *4.2 Challenges which might emerge in single-molecule biophysics*

Here we discuss some putative challenges relating to future single-molecule biophysics techniques, based on interpolating our current knowledge.

*4.2.1 Challenges in statistics*
Single-molecule biophysics approaches are in general inherently low throughput, and examining enough molecules to reach a statistically significant result can be extremely time consuming. There are a number of emerging strategies to increase molecular throughput, either by increasing the number of molecules observed at any one time, or by increasing efficiency of observing molecules sequentially. The drive to higher throughput super-resolution fluorescence microscopy is in several cases enabled by new microfluidics. Microfluidics enable manipulation of fluid flow to move single particles or change fluid conditions within a sample chamber. For example, in DNA curtains liquid flow over micro-fabricated substrates is used to produce multiple aligned DNA strands used to study the interactions of proteins with DNA, with multiple proteins recorded in a single field of view. Alternatively, narrow channels are used, with the single molecules studied moved through sequentially, allowing them to be fluorescently imaged or manipulated (e.g. using OT), with less time between repeat experiments than in earlier experiments where a researcher needed to locate individual particles and/or wait for them to diffuse into view.

OT can be parallelised using a range of methods: acousto-optical deflectors (AODs), scanning mirrors and interference can produce multiple traps via laser beam time-sharing. Holography uses spatial light modulators to create multiple OT in the same sample chamber and can create higher numbers of parallel traps than time-sharing approaches. Multiple laser beams can also be used to decrease acquisition times in super-resolution microscopy, e.g. with multiple STED 'donuts' to acquire images faster (25).

It is clear that high throughput approaches are important for bringing single-molecule biophysics to real world problems, such as medical diagnosis, but not all the techniques are currently available in high throughput versions.

*4.2.2 Checking assumptions*
The first single-molecule biophysics experiments required several simplifying assumptions to evaluate parameters. As complexity of experimental designs increased, we have moved further from a realm in which such assumptions are universally correct and must be mindful of the assumptions on which analysis techniques are based. E.g. In most applications of fluorescence microscopy, it is a basic assumption that the fluorescence dipole is free to rotate and produces a symmetric image during the time for a single sampling time window. However, when a dye molecule is bound to a biomolecule of interest it can become highly orientated, for example when bound to immobilised DNA, producing a non-symmetric image and resulting in an error-prone localisation estimate. This is a concern when localisation precisions <10 nm are found and should be accounted for in experiments where the rotation of the fluorescence dipole is constrained.

Another example is in STORM analysis software. These programs typically assume that each bright fluorescent focus is a single fluorophore. As technological advances have driven higher-throughput STORM imaging it is now often the case that higher numbers of molecules are present within a single diffraction-limited PSF width, and care must be taken to ensure the software used can recognise this to avoid mis-localisation.

Such issues are not limited to optical microscopy: in electron microscopy the alignment and clustering algorithms are sometimes found to produce outcomes that strongly resemble initial guesses. Care must be taken to compare results to initial parameterisation conditions, or new methods that check for this must be used, a challenge which is increasingly evident with AI approaches are used since the outputs are steered by the quality and nature of the training datasets employed and can potentially result in so-called "hallucinations".

The huge expansion in all areas of single-molecule biophysics has allowed experimental complexity to increase in the last ten years, but amongst this progress it is important to continue to check underlying assumptions of our methods as we move to higher temporal resolution and studying biological systems with increased dynamic complexity.

*4.2.3 Big Data Management*

Approaches to manage Big Data images involve every aspect of data analysis. Good practice to keep compatibility between metadata, file formats, and open data interfaces is invaluable. Cloud storage and processing facilities can resolve storing, sharing and intense processing problems that are too challenging for local machines. Automated data selection and analysis approaches are increasingly valuable to keep pace with the huge volumes of data that are now being generated, not atypically ~1 TB per experiment.

*4.2.4 Challenges of an increasingly non-specialist userbase*

The last decade has seen substantial growth in the number of single-molecule biophysics techniques, and efforts has been made to develop protocols for making samples, characterise the machines and their limits and develop analysis software. Much of this work has been carried out on test sample systems that are already well characterised. For example, microtubule imaging (Fig. 11) has become one of the gold standards in super-resolution microscopy in having a well-defined diameter of 24 nm which is in the ballpark for good super-resolution spatial precision estimates.

The development of new super-resolution approaches is making a significant impact to the complexity of single-molecule biophysics experiments of live organisms and the associated increase in physiological relevance from the results that emerge. Nevertheless, the new experimental instrumentation becomes ever more challenging to operate, in particular for research users who increasingly come from a non-specialist background in regard to bespoke optical microscopy and physics, such as cell biologists. There have been efforts for light sheet microscopy with the OpenSPIM project to increase to usability of instrumentation and democratise access to it away from optical physics laboratories, which has resulted in a high uptake of the technique amongst biologists. Also, there are now several commercial systems that STORM and related PALM type imaging, but they are not in general as simple to customise as bespoke systems.

Combined with the increased use of single-molecule biophysics technologies by non-specialists there are now several open-source data analysis software packages available particularly for super-resolution microscopy. Although such software packages typically come associated with comprehensive manuals of operation it will become increasingly important for users to try to engage with the core physical principles on which techniques work to optimise imaging conditions and avoid artefacts etc. Reviews of approaches aimed at non-specialists serve to address this challenge, but still greater communication between developers and users is likely to be helpful.

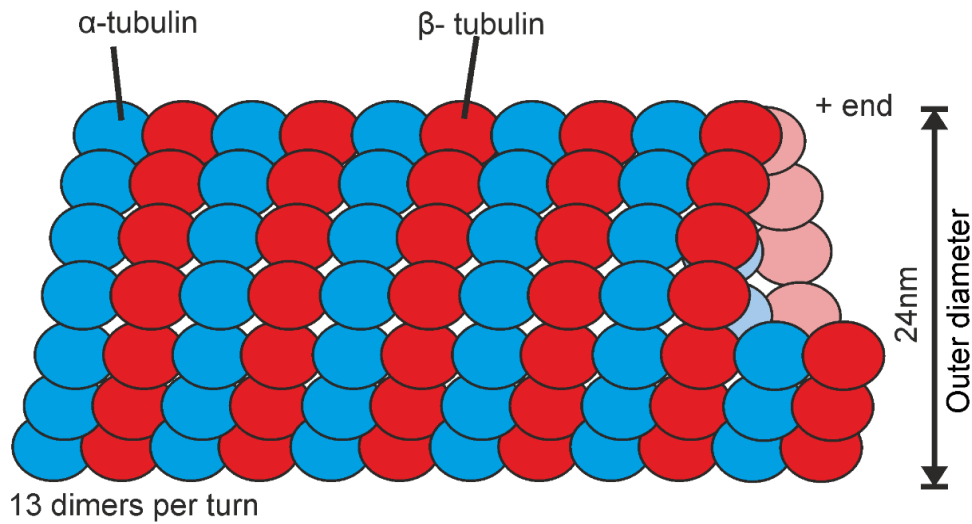

*Figure 11: Cartoon of microtubule structure. Repeating tubulin dimers can be labelled with fluorescent dyes to provide a regular structure for testing the localisation precision capabilities of new super-resolution microscopy techniques.*

*4.2.4 Single-molecule biophysics in populations of cells*

There are a range of challenges for imaging single molecules inside populations of living cells, such as in tissues. E.g., the noise due to scattering in often inhomogeneous and optically deep tissue environments, the biocompatibility of fluorescent dyes in tissues, the number of emitted photons from dyes and their fluorescence lifetime in a tissue environment. Emerging approaches such as *in vivo* FRET and fluorescence imaging using NIR light tackle these issues to some extent. Adaptive optics have proved valuable in correcting for the inhomogeneity in refractive index in a deep tissue sample. When used in combination with selective plane illumination microscopy (SPIM) which uses transverse illumination, a light sheet, this can substantially reduce the level of out of focal plane scattered light from deep tissue samples, to enable single-molecule detection in multicellular tissues up to a few tenths of a mm deep.

*4.2.5 Moving towards personalized medicine*

Personalized medicine aims to cater healthcare specifically to an individual patient and so targets more effective diagnosis and treatment with fewer side effects. There has been recent progress to understand the biophysics of infection (26–29), with the relevance to single-molecule biophysics principally from miniaturized biosensing and LOC devices to enable smart diagnostics of health disorders, but also in the application of targeted treatment and cell delivery tool such as those of "nanomedicine". Developments in microfluidics including surface chemistry methods, nanophotonics and bioelectronics have facilitated miniaturization of biosensing instrumentation to a level of single biomolecules sensitivity. These tools typically comprise silicon-based substrates in microscopic flow cells for detecting specific biomolecules, with synthetic arrangements of biological material bound to surfaces inside the flow cell, in a complex microfluidics design to convey dissolved sample from a range of biological fluids such as blood, urine, and sputum etc., to detection zones inside the device. For detection of specific biomarkers, which label specific biomolecules, a surface pull-down approach is often used in which the surface within a detection zone on the chip is functionalised to bind specifically to one or more biomarkers.

Surface-immobilized the pulled down molecule can then be detected using several different biophysics approaches. Commonly, fluorescence detection is used applied if the biomarker can be fluorescently tagged, enabled though photonics properties of the silicon-based flow cell such as waveguiding of excitation light to different regions of the device, and photonic bandgap filtering used to separate excitation from emission wavelengths. Microfabricated photonic surface geometries can also be engineered to generate evanescent excitation fields to increase the detection signal-to-noise- ratio by minimizing signal detection from unbound biomolecules away from the surface.

Label-free detection LOC is also emerging, including interferometry surface plasmon resonance (SPR) methods, Raman spectroscopy, electrical impedance and quartz crystal microbalance (QCM) detectors that operate through detecting changes in resonance frequency due to the surface binding of biomolecules.  Cantilevers similar to those in AFM can be used for single biomolecule detection, involving chemical functionalization of the cantilever, which results in binding of a specific biomolecules as a sample solution is flowed across, detected as a change in cantilever resonance frequency.

Substantial recent developments which work towards personalised medicine have been made in efficiently and cheaply sequencing single molecules of DNA, as discussed earlier in this review, involving ion conductance measurement through nanopores. Also, the increasing use of nanomedicine is at the level of targeted drug binding is already emerging. E.g. pharmaceutical treatments which destroy specific cells such as cancer cancers, using radioactive nanoparticles coated with specific single-molecule antibody probes. Aptamers (synthetic engineered recognition molecules composed either of nucleic acids or more rarely peptides) play a potentially valuable role in stimulating minimal immune response compared to antibodies, thus resulting in fewer side-effects. Targeted binding is also showing promise for visualising disease in tissues, such as using antibody-tagged quantum dots to specifically bind to cancer tumours.

Targeted drug delivery tools which increase the specificity and efficacy of drugs being internalized by the cells is a popular area of new research. Systems are being designed which use modify the normal process of endocytosis by which many cells internalize biomolecules, but also other strategies which use DNA origami devices to form synthetic 3D nanostructures made from DNA to acts as molecular cages that encapsulate different drugs deep to protect them from degradation before they can be released at this intended point of action.

A promising areas of nanomedicine research is tissue regeneration using 'biomimetic' materials, which is benefiting from single-molecule biophysics methods regarding their physical characterisation. These materials can serve as replacements for damaged or diseased tissues to permit stem cells to assemble in highly specific regions of space to facilitate growth of new tissue. They focus on mimicking key structural properties of healthy tissue, e.g. bone and teeth but also softer structural material such as collagen but also use inorganic biomimetics that exhibit chemical and structural stability, such as noble metals, metal oxide semiconductors, chemically inert plastics and ceramics, often involving pre-coating with short sequence peptides to encourage binding of cells from surrounding tissues.

## 5. Conclusions

It is clear that single-molecule biophysics has outgrown the earlier constraints of the pioneering techniques of structural biology and physiology, establishing a new path forward in moving towards approaches which render single-molecule precise data while still retaining physiological function of the biological system under study. A key feature with the development of single-molecule biophysics has been that the biological questions often come before the technology development; in other words, the biology problem necessitates the physics solution. But what is also clear is that these single-molecule biophysics studies can then motivate the development of new physics; for example, understanding how single-molecule physics actually relatives to ensemble level statistical thermodynamics theory. This has been demonstrated in understanding how single-molecule transitions of molecular machines can relate free energy differences between two states (such as being folded and unfolded) and the irreversible work done along the ensemble of trajectories joining the same states; the theory is embodied in the Jarzynski equality, but it is only using single-molecule biophysics tools such as OT and AFM that it was seen the free energy landscape for such transitions was more granular than that predicted, for example involving multiple pathways and states (30). Also, single-molecule biophysics studies at molecular scales, especially using imaging approaches on relatively large tissue sections, is beginning to unpick how the physical rules of interaction change during emergence of new tissue-level properties from the underpinning molecular agents.

A drive for searching for simplicity for any new physics remains however; this contrasts with traditional biology modelling which largely strives to expand on the number of parameters used to account for new and more complex observations of a given biological process, whereas the new physics approach still seeks to explain increased complexity of any process through simplification to the barest suite of non-redundant physical features – a Holy Grail quest, if you like, for the fewest number of scaling parameters. An over-arching new physics theme which single-molecule biophysics is contributing towards is the development of new theories explicitly inspired by the cross-scale nature of biology. For example, this can be seen in the development of new theory that accounts for phase transitions of biological matter. Phase transitions are ubiquitous in the known Universe, seen in the familiar transitions of water to ice and steam through to extraordinary astrophysical behaviour of the coalescence of stars and galaxies. However, recent insights from biomolecular liquid-liquid phase separation (LLPS), driven in part through single-molecule biophysics approaches, reveal preferences for length scales within the transitions which have not been predicted by existing theory that primarily is based on traditional polymer physics mixture theory (31,32).

In biomolecular LLPS, biomolecules, usually proteins and RNA, partition from their water-solvated states into mesoscale liquid droplets inside cells in which water is excluded. These droplets have interesting properties in enabling partitioning of specific chemical reactions to small volumes inside a cell and hence potentially increasing the chemical efficiency significantly due to a very high local enhancement of concentration of reactants. Traditional theory predicts that such phase transitions ultimately go to completion to utilise all the molecular components involved in a given cell, however, what is actually observed in steady state is a preference for a range of droplet diameters. These experimental details have been revealed primarily through advances in rapid single-molecule biophysics imaging tools and the new physics rules explaining this behaviour likely result from cross-scale effects in how multiple molecular interactions forces in the droplet, mediated through van der Waals and electrostatic forces, integrate to affect higher length and time scale physical properties such as droplet viscosity and surface tension, which ultimately imposes limits to droplet diameter.

If pedantic, one might argue that such phenomena are not "new" physics, however, I am inclined to suggest that this reasoning is largely erroneous semantics; new "physical rules" which emerge due to feedback of physical outputs across multiple scales in a system does constitute a new type of physics. This general cross-scale new physics theme might be best denoted as "systems biophysics", a form of physics of emergence which is inspired by biology, and which can also facilitate engineering the development of new biomaterials based on cross-scale physics. Single-molecule biophysics has in effect come full circle in this regard, in that the demands of developing new physics understanding of biology mean that we require single-molecule details but not in isolation as single molecules but rather embedded in the context of the complex cross-scale milieu that is real biology.

**Glossary of acronyms**

| Acronym | Definition |
|---|---|
| AFM | Atomic force microscopy |
| AI | Artificial intelligence |
| AT | Acoustic tweezers |
| ATP | Adenosine triphosphate |
| BaLM | Bleaching/Blinking assisted Localization Microscopy |
| BALM | Binding-Activated Localization Microscopy |
| BSE | Bovine serum albumin |
| CCD | Charge-coupled detector |
| CLEM | Correlative light and electron microscopy |
| COMBI-Tweez | Combined Optical and Magnetic BIomolecule TWEEZers |
| Cryo-EM | Cryogenic electron microscopy |
| DNA | Deoxyribose nucleic acid |
| EFM | Electrostatic force microscopy |
| EM | Electron microscopy |
| ExM | Expansion microscopy |
| FCS | Fluorescence correlation microscopy |
| FLIP | Fluorescence loss in photobleaching |
| FRAP | Fluorescence recovery after photobleaching |
| FRET | Förster resonance energy transfer |
| HOT | Holographic optical tweezers |
| HS-AFM | High speed atomic force microscopy |
| iSIM | Instant structure illumination microscopy |
| LG | Laguerre-Gaussian |
| LLPS | Liquid-liquid phase separation |
| LOC | Lab on a chip |
| MINFLUX | Minimal fluorescence photon fluxes |
| MT | Magnetic tweezers |
| NA | Numerical aperture |
| NIR | Near infrared |
| OpenSPIM | Open selective plane illumination microscopy |
| OT | Optical tweezers |
| PAINT | Points accumulation for imaging in nanoscale topography |
| PALM | Photoactivated localisation microscopy |
| PSF | Point spread function |
| QCM | Quartz crystal microbalance |
| QPD | Quadrant photodiode |
| RESOLFT | Reversible saturable optical linear fluorescence transitions |
| RNA | Ribonucleic acid |
| SICM | Surface ion conductance microscopy |
| SIM | Structured illumination microscopy |
| SlimVar | Variable-angle Slimfield microscopy |
| SLM | Spatial light modulator |
| smFRET | Single-molecule Förster resonance energy transfer |
| SPIM | Selective plane illumination microscopy |
| SPM | Surface probe microscopy |
| SPR | Surface plasmon resonance |
| SSIM | Saturated structured illumination microscopy |
| STED | Stimulation emission depletion |
| STORM | Stochastic optical reconstruction microscopy |
| TIRF | Total internal reflection fluorescence |
| UEM | Ultrafast electron microscopy |

Acknowledgements

This work was supported by the Engineering and Physical Sciences Research Council EPSRC (grant EP/Y000501/1).